\documentclass[epj]{svjour}
\usepackage{graphics}
\begin{document}
\title{Distinct ground state features and the decay chains of Z = 121 Superheavy Nuclei}
\author{G. Saxena\inst{1,2} \and U. K. Singh\inst{2}\and M. Kumawat\inst{1,2}\and M. Kaushik\inst{3} \and S. K. Jain\inst{2}\and Mamta Aggarwal\inst{4}}
\institute{Department of Physics, Govt. Women Engineering College,
Ajmer-305002, India \and Department of Physics, School of Basic Sciences, Manipal
University, Jaipur-303007, India\and Department of Physics, Shankara Institute of Technology, Kukas, Jaipur-302028, India\and Department of Physics, University of Mumbai, Kalina,
Mumbai-400098, India}
\date{Received: date / Revised version: date}
\abstract{
A fully systematic study of even and odd isotopes (281 $\leq$ A $\leq$ 380) of Z = 121 superheavy nuclei is presented in theoretical frameworks of  Relativistic mean-field plus state dependent BCS approach and Macroscopic-Microscopic approach with triaxially deformed Nilson Strutinsky prescription. The ground state properties namely shell correction, binding energy, two- and one- proton and neutron separation energy, shape, deformation, density profile and the radius are estimated that show strong evidences for magicity in N = 164, 228. Central depletion in the charge density due to large repulsive Coulomb field indicating bubble like structure is reported. A comprehensive analysis for the possible decay modes specifically $\alpha$-decay and spontaneous fission (SF) is presented and the probable $\alpha$-decay chains are evaluated. Results are compared with FRDM calculations and the available experimental data which show excellent agreement.
}
\maketitle
\section{Introduction}
\label{intro}
To search for the possible fusion reactions and to identify potential superheavy nuclei especially the ones with Z $>$ 118 is one of the eminent problems in the current nuclear physics world. Up to Z = 118, many superheavy nuclei have been produced either by cold fusion reaction with target $^{208}$Pb and $^{209}$Bi at GSI \cite{Hofmann2000,Hofmann2011} and RIKEN \cite{Morita2007} or by hot fusion with projectile $^{48}$Ca at JINR \cite{Oganessian2010,Oganessian2015,Hamilton2013}. For higher Z = 119, 120, few attempts \cite{Hofmann2016,Oganessian2009} have already been made towards exploring the superheavy nuclei. However the scope to explore further and search for distinct features beyond Z $>$ 120 is limitless.\par
Decay properties of various superheavy nuclei \cite{Buck1992,Poenaru1984,Basu2002,Zhang2007,Sharma2004,Pei2007} and few possible new shell closures in this unknown territory of superheavy region with Z = 120, 132, 138 and N = 172, 184, 198, 228, 238, 258 \cite{Moller1994,Rutz1996,Cwiok1996,Zhang2004} have been reported theoretically. Some ground state properties and decay modes have been studied \cite{Bao2015,Wang2015,Niyti2015,Heenen2015,Santhosh2016,Budaca2016,Liu2017,Zhang2017} for Z = 119, 120 \cite{Poenaru2017}, Z = 121 \cite{Santhosh2016}, Z = 122 \cite{Manjunatha2016}, Z = 120, 124 \cite{Mehta2015}, Z = 123 \cite{Santhosh2016a}, Z = 122, 124, 126, 128 \cite{Santhosh2016xrd}, Z = 124, 126 \cite{Manjunatha2016bbm,Manjunatha2016zia}, Z = 132, 138 \cite{Rather2016} and the search for new shell closures in this region \cite{Wu2003,Zhang2004,Adamian2009,Biswal2014} has become a thrust area of research in the recent times.  The less explored phenomenon of bubble/semi-bubble structures in superheavy region has also been reported in few recent works \cite{Berger2001,Decharge2003,Afanasjev2005,SinghSK2013,Saxenaplb2018}. \par
In view of the growing interest and curiosity in this new domain of superheavy nuclei, we move one step closer to the experimental attempts \cite{Hofmann2016,Oganessian2009} and investigate even and odd isotopes of Z = 121 (293 $\leq$ A $\leq$ 380) using two simple and effective well established theoretical formalisms (i) Relativistic mean-field plus state dependent BCS (RMF+BCS) approach \cite{Serot1984,Ring1996,Yadav2004,Saxena2017,Saxena2017hzo} and (ii) Macroscopic-Microscopic approach with triaxially deformed Nilson Strutinsky method (NSM) \cite{Aggarwal2010,Aggarwal2014}. In this work, we aim to dig into (i) the ground state properties namely the shell correction, binding energy, separation energy, deformation, shape, charge density and neutron density and the radius of the lesser known Z = 121 superheavy nuclei (ii) the phenomenon of proton bubble/semi-bubble formation, if any (iii) possible $\alpha$-decay chains for the identification of new elements. Our calculations provide significant inputs about the ground state properties and magicity in N = 164 and 228, existence of bubble structure and identification of $\alpha$-decay chains and the decay modes in Z = 121 isotopes. We compare our results with the available experimental data \cite{Wang-Mass2017} and Finite Range Droplet Model (FRDM) calculations \cite{Moller2012} which show good agreement and may be useful to guide future experiments to explore Z = 121 superheavy nuclei.

\section{Theoretical Formalisms}
\subsection{Relativistic Mean-Field Theory} RMF calculations have been carried out using the model Lagrangian density with nonlinear terms both for the ${\sigma}$ and ${\omega}$ mesons as described in detail in Refs. \cite{Singh2013,Yadav2004}.
\begin{eqnarray}
       {\cal L}& = &{\bar\psi} [\imath \gamma^{\mu}\partial_{\mu}
                  - M]\psi\nonumber\\
                  &&+ \frac{1}{2}\, \partial_{\mu}\sigma\partial^{\mu}\sigma
                - \frac{1}{2}m_{\sigma}^{2}\sigma^2- \frac{1}{3}g_{2}\sigma
                 ^{3} - \frac{1}{4}g_{3}\sigma^{4} -g_{\sigma}
                {\bar\psi}  \sigma  \psi\nonumber\\
               &&-\frac{1}{4}H_{\mu \nu}H^{\mu \nu} + \frac{1}{2}m_{\omega}
                  ^{2}\omega_{\mu}\omega^{\mu} + \frac{1}{4} c_{3}
                 (\omega_{\mu} \omega^{\mu})^{2}
                  - g_{\omega}{\bar\psi} \gamma^{\mu}\psi
                 \omega_{\mu}\nonumber\\
              &&-\frac{1}{4}G_{\mu \nu}^{a}G^{a\mu \nu}
                 + \frac{1}{2}m_{\rho}
                 ^{2}\rho_{\mu}^{a}\rho^{a\mu}
                  - g_{\rho}{\bar\psi} \gamma_{\mu}\tau^{a}\psi
                 \rho^{\mu a}\nonumber\nonumber\\
               &&-\frac{1}{4}F_{\mu \nu}F^{\mu \nu}
                 - e{\bar\psi} \gamma_{\mu} \frac{(1-\tau_{3})}
                 {2} A^{\mu} \psi\,\,,%\nonumber\
\end{eqnarray}
where the field tensors $H$, $G$ and $F$ for the vector fields are
defined by
\begin{eqnarray}
                 H_{\mu \nu} &=& \partial_{\mu} \omega_{\nu} -
                       \partial_{\nu} \omega_{\mu}\nonumber\\
                 G_{\mu \nu}^{a} &=& \partial_{\mu} \rho_{\nu}^{a} -In t
                       \partial_{\nu} \rho_{\mu}^{a}
                     -2 g_{\rho}\,\epsilon^{abc} \rho_{\mu}^{b}
                    \rho_{\nu}^{c} \nonumber\\
                  F_{\mu \nu} &=& \partial_{\mu} A_{\nu} -
                       \partial_{\nu} A_{\mu}\,\,\nonumber\
\end{eqnarray}
and other symbols have their usual meaning. The corresponding
Dirac equations for nucleons and Klein-Gordon equations for mesons
obtained with the mean-field approximation are solved by the
expansion method on the widely used axially deformed
Harmonic-Oscillator basis \cite{Geng2003,Gambhir1989}. The quadrupole
constrained calculations have been performed for all the nuclei
considered here in order to obtain their potential energy surfaces
(PESs) and determine the corresponding ground-state deformations
\cite{Geng2003,Flocard1973}. For nuclei with odd number of nucleons,
a simple blocking method without breaking the time-reversal
symmetry is adopted \cite{Geng2003wt,Ring1996}.

In the calculations we use for the pairing interaction a delta force, i.e., V = -V$_0 \delta(r)$ with the strength V$_0$ = 350 MeV fm$^3$ which has been used in Refs. \cite{Yadav2004,Saxena2017} for the successful description of drip-line nuclei. Apart from its simplicity, the applicability and justification of using such a $\delta$-function form of interaction has been discussed in Ref. \cite{Dobaczewski1983}, whereby it has been shown in the context of HFB calculations that the use of a delta force in a finite space simulates the effect of finite range interaction in a phenomenological manner (see also \cite{Bertsch1991} for more details).

Whenever the zero-range $\delta$ force is used either in the BCS
or the Bogoliubov framework, a cutoff procedure must be applied,
i.e. the space of the single-particle states where the pairing
interaction is active must be truncated. This is not only to
simplify the numerical calculation but also to simulate the
finite-range (more precisely, long-range) nature of the pairing
interaction in a phenomenological way \cite{Dobaczewski1995,Goriely2002}. In
the present work, the single-particle states subject to the
pairing interaction are confined to the region satisfying
\begin{equation}
\epsilon_i-\lambda\le E_\mathrm{cut},
 \end{equation}
 where $\epsilon_i$ is the single-particle energy, $\lambda$
 the Fermi energy, and $E_\mathrm{cut} = 8.0$ MeV. The center-of-mass correction is approximated by
 \begin{equation}
 E_{\textrm{cm}} = -\frac{3}{4}41A^{-1/3},
 \end{equation}
which is often used in the relativistic mean field theory among
the many recipes for the center-of-mass correction
\cite{Bender1999}. For further details of these formulations we refer the
reader to Refs. \cite{Singh2013,Geng2003,Gambhir1989}. In the next section, we give a brief description of Macroscopic-Microscopic approach.

\subsection{Macroscopic - Macroscopic approach using Nilsson-Strutinsky method (NSM)}
Macroscopic-Microscopic approach using the traixially deformed Nilsson Strutinsky method (NSM) treats the structural properties of the atomic nuclei which are governed by the delicate interplay of macroscopic bulk properties of the nuclear matter and the microscopic shell effects which start with the well known Strutinsky density distribution function \cite{Strutinsky1968,BRACK1972,Nilsson1972} for single particle states
\begin{equation}
\tilde {g(\epsilon)} = {1 \over {\sqrt{\pi}\gamma}} \sum exp(-u_i)^2 \sum_{k = 0 }^{\infty} C_k H_k (u_i),
\end{equation}
where
\begin{equation}
u_i = {{(\epsilon - \epsilon_i)} \over \gamma},
\end{equation}
and the coefficients C$_k$ are
\begin{equation}
C_k =
{\begin{array}{ccc} \frac{{-1}^{k/2}}{{2^k}{{(k/2)}!}} & k  & even \\
0 & k  & odd
\end{array}}
\end{equation}
Hermite polynomials H$_k$(u$_i$) upto higher order of correction ensures smoothened levels. The energy due to Strutinsky's smoothed single particle level distribution is given by
\begin{equation}
\tilde E = \int_{-\infty}^{\mu} \tilde g(\epsilon) d\epsilon
\end{equation}
The chemical potential $\mu$ is fixed by the number conserving equation
\begin{equation}
N = \int_{-\infty}^{\mu} \tilde g(\epsilon) d\epsilon
\end{equation}
The shell correction to the energy is obtained as usual
\begin{equation}
\delta E_{Shell} = \sum_{i=1} ^A \epsilon_i- \tilde E
\end{equation}
where the the smearing width of 1.2$\hbar$$\omega$ has been used. The single particle energies $\epsilon$$_i$ as a function of  deformation parameters ($\beta$, $\gamma$) are generated by Nilsson Hamiltonian for the triaxially deformed oscillator diagonalized in a cylindrical representation \cite{Shanmugam1979,EIS1976}.
\begin{eqnarray}
H = p^2 /2m + (m/2)(\omega_x^2 x^2 + \omega_y^2 y^2 +\omega_z^2 z^2)+ \nonumber\\
 C l.s + D (l^2- 2<l^2>)
\end{eqnarray}
The coefficients for the l.s and l$^2$ terms are taken from Seeger \cite{Seeger1975} who has fitted them to reproduce the shell corrections \cite{Strutinsky1968} to ground state masses. Strutinsky's  shell correction $\delta$E$_{Shell}$ added to macroscopic energy of the spherical drop BE$_{LDM}$ along with the deformation energy E$_{def}$ obtained from surface and Coulomb effects gives the total energy  BE$_{gs}$ as in our earlier works \cite{Aggarwal2010,Aggarwal2014} corrected for microscopic effects of the nuclear system
\begin{eqnarray}
BE_{gs}(Z,N,\beta,\gamma) = BE_{LDM}(Z,N)-E_{def}(Z,N,\beta,\gamma) \nonumber\\
-\delta E_{shell}(Z,N,\beta,\gamma)\,\,\,\,\,\,\,\,\,\,\
\end{eqnarray}
Energy E (= -BE) minima are searched for various $\beta$ (0 to 0.4 in steps of 0.01) and $\gamma$ (from -180$^o$(oblate) to -120$^o$(prolate) and -180$^o$ $<$ $\gamma$ $<$ -120$^o$ (triaxial)) to trace the equillibrium deformations and nuclear shapes respectively. The first (1p and 2p) and (1n and 2n) unbound nuclei are located by one and two proton separation energy and neutron separation enetgy approaching zero value obtained as the difference between the binding energies BE$_{gs}$ of the parent and daughter nucleus.

\section{Results and discussions}
The results of the present work are two fold:\par (i) Study of ground state properties specifically binding energies, separation energies, shell correction, deformation, shape, radii and charge density of Z = 121 superheavy nuclei (A = 281 $-$ 380) and search of new magic and bubble like structures, and \par
(ii) Investigation of possible decay modes such as $\alpha$-decay and spontaneous fission (SF) and to identify $\alpha$-decay chains.
\subsection{Ground State Properties}
\begin{figure}
\resizebox{0.5\textwidth}{!}{%
  \includegraphics{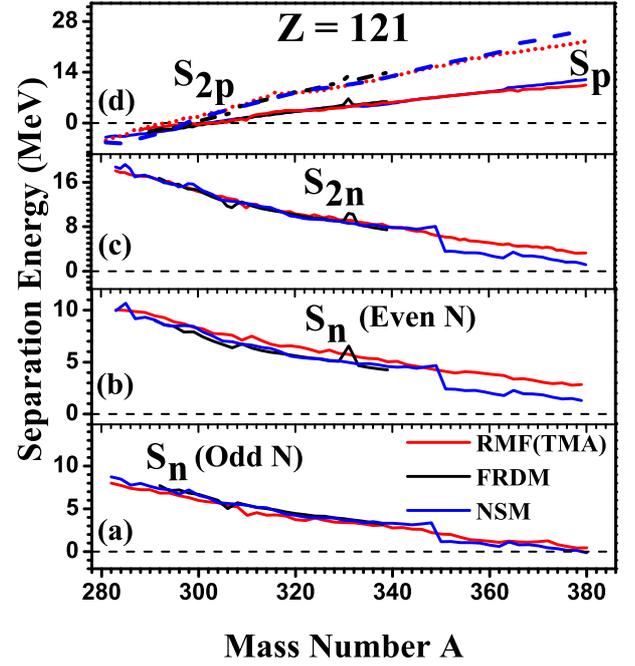}
}
%\vspace{5cm}       % Give the correct figure height in cm
\caption{(Colour online) (a) S$_{n}$ (for odd N), (b)  S$_{n}$ (for even N) and (c) S$_{2n}$, (d) S$_{p}$ and S$_{2p}$ of Z = 121 isotopes vs. A.}
\label{fig1}
\end{figure}
Fig. \ref{fig1} shows the complete trace of one- and two- neutron separation energy (S$_{n}$ and S$_{2n}$) and one- and two- proton separation energy (S$_{p}$ and S$_{2p}$) for Z = 121 isotopes with A = 281 $-$ 380 evaluated using RMF approach \cite{Yadav2004,Saxena2017} and Mac-Mic approach with NS prescription (NSM) \cite{Aggarwal2010,Aggarwal2014}. FRDM \cite{Moller2012} data values available so far are plotted for comparison. It may be noted that all the three approaches (RMF, NSM and FRDM) are showing agreement in this superheavy domain and validate our calculations. The first one- and two-proton unbound nuclei in Z = 121 superheavy isotopic chain are predicted to be $_{183}^{304}121$, $_{180}^{301}121$ and $_{171}^{292}121$, $_{176}^{297}121$ from RMF and NSM calculations respectively. At A = 350,  we note a sharp drop in NSM values of S$_{2n}$ and S$_{n}$ which indicate a shell closure at A = 349 (Z $=$ 121, N $=$ 228) and N $=$ 228 emerges as a new neutron magic number which was speculated by other theoretical works \cite{Zhang2004,Lombard1976,Patra1999} also. Another probable candidate for magicity is N $=$ 164 (A = 285) that lies  beyond the proton drip line with negative separation energy. At N $=$ 164, the separation energy shows a maxima and then drops to a lower value with increasing N $=$ 165  indicating a shell closure.\par
\begin{figure}
\resizebox{0.5\textwidth}{!}{%
  \includegraphics{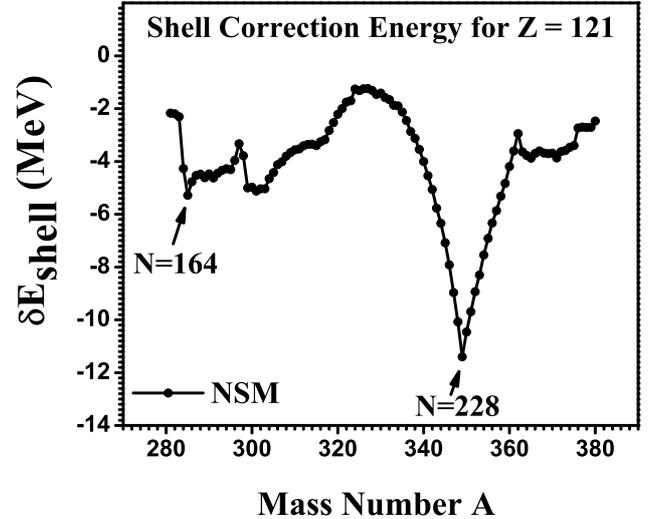}
}
\caption{Variation of Shell correction $\delta$E$_{Shell}$ (in MeV) vs. A. The deep minima indicates magicity at N = 228 and 164.}
\label{fig2}
\end{figure}
The magicity in  N $=$ 164 and 228 is also evident in Fig. \ref{fig2} which shows our estimate of shell correction values $\delta$E$_{Shell}$ (using Eq (9)) evaluated for A $=$ 281 $-$ 380 isotopes of Z = 121. The shell correction to energy $\delta$E$_{Shell}$ is expected to show a minima at around shell closures. In Fig. \ref{fig2}, we note  a very deep minima of around 12 MeV in $\delta$E$_{Shell}$ value at  N = 228 which shows the strong magic character. Another shell closure is located at N $=$ 164 with a relatively shallower minima of around 5.28 MeV. Since the magic number nuclei are expected to have zero deformation with spherical shape, it is important to explore structural transitions for a conclusive viewpoint on magicity in Z $=$ 121 superheavy region. \par
\begin{figure}
\resizebox{0.5\textwidth}{!}{%
  \includegraphics{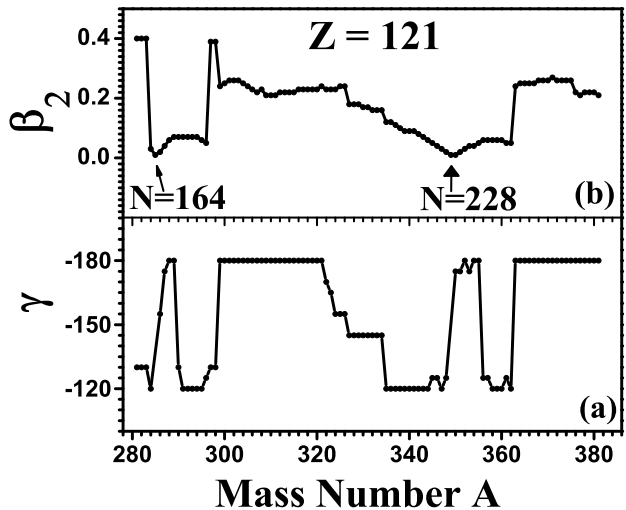}
}
\caption{Variation of $\beta_2$ and $\gamma$ vs. A for Z = 121 (using NSM) where the shapes are denoted by $\gamma$ = -180$^o$ (oblate), -120$^o$ (prolate) and all other (triaxial).}
\label{fig3}
\end{figure}
For the study of nuclear shapes, the most closely related experimental observables are the quadrupole moments of excited states and electromagnetic transition rates and their measurements \cite{Wood1992,Julin2001} that provide impetus to test the predictions of the  theoretical models of nuclear structure. However the superheavy region under consideration here has not been explored much and the experimental data is yet scarce, hence we use theoretical model to locate energy minima with respect to Nilsson deformation parametrs ($\beta$, $\gamma$) and present for the first time a complete trace of equilibrium deformations and shapes along the whole chain of Z $=$ 121 isotopes with A $=$ 281 $-$ 380  in  Fig. \ref{fig3}. Nuclei in this region are found to be well deformed with $\beta$ ranging between 0.2 - 0.4 with two minima of zero deformation at A $=$ 285 (N $=$ 164) and A $=$ 349 (N $=$ 228) which indicate shell closure. \par
\begin{figure}
\resizebox{0.5\textwidth}{!}{%
  \includegraphics{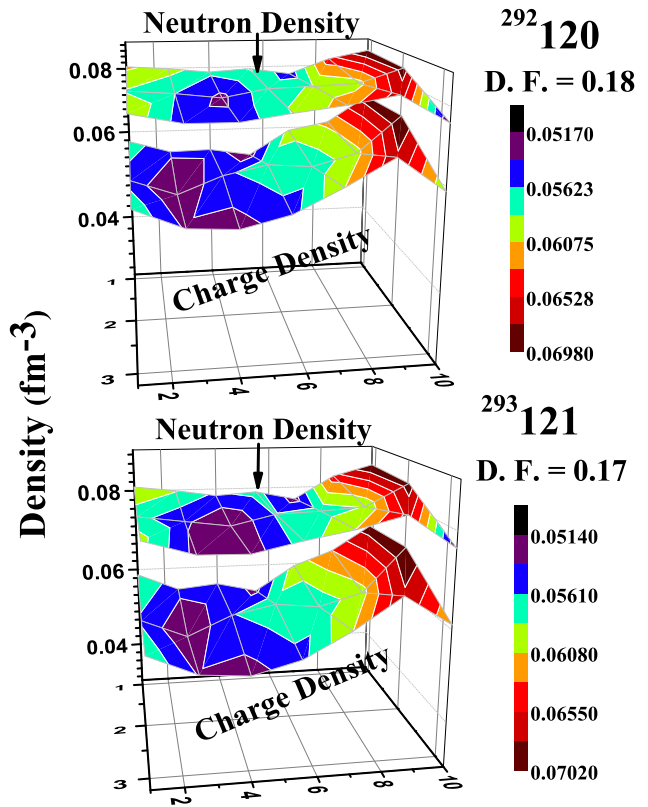}
}
\caption{(Colour online) Radial variation of charge density and neutron density within RMF+BCS approach for $^{292}$120 and $^{293}$121.}
\label{fig4}
\end{figure}
The evolution of nuclear shapes is shown in Fig. \ref{fig3}(a) where we plot shape parameter $\gamma$ vs A. A series of shape transitions from oblate ($\gamma$ $=$ -180$^o$) at (A = 299 - 321) to triaxial (-120$^o$ $<$ $\gamma$ $<$ -180$^o$)  at A $=$ 322 - 334 to few prolate ($\gamma$ $=$ -120$^o$) at A $=$ 335 - 348 is observed. The spherical shape with zero deformation at N $=$ 164, 228 (A $=$ 281, 349) shows magic character. While undergoing shape transitions, many nuclei appear to be potential candidates for shape coexistence which will be discussed in our upcoming work. Shell correction to energy $\delta$E$_{shell}$ (see Fig. \ref{fig2}) varies from $\approx$ -12 MeV (at N = 228) upto 1 MeV (at mid shell) which points towards the shape transitions from spherical (at N $=$ 228) to well deformed region (mid shell) which is evident in the plots of deformation and shapes (Figs. \ref{fig3} (a) and (b)) and reaffirms the magic character in N $=$ 164 and 228. \par
\begin{figure}
\resizebox{0.5\textwidth}{!}{%
  \includegraphics{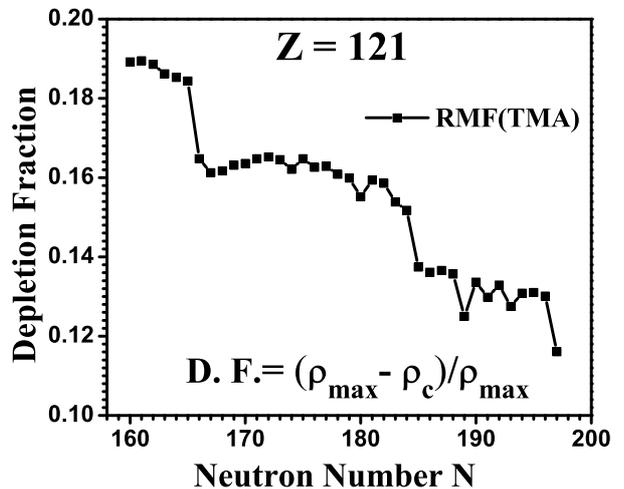}
}
\caption{Variation of depletion fraction D.F. with N for Z = 121 isotopes.}
\label{fig5}
\end{figure}
Based on some recent studies on the phenomenon of bubble structure \cite{Berger2001,Decharge2003,Afanasjev2005,SinghSK2013,Saxenaplb2018,Li2016,Schuetrumpf2017,Duguet2017}, the major cause of central depletion in the charge density is found to be either due to depopulation of s state or due to large repulsive Coulomb field. The unoccupied s-state has been studied theoretically \cite{Li2016,Schuetrumpf2017,Duguet2017} and experimentally in sd-shell nuclei \cite{Mutschler2016} whereas the large repulsive Coulomb field is expected to occur dominantly in superheavy nuclei due to large number of protons \cite{Saxenaplb2018}. With this in view, we investigate depletion in central density (bubble/semi-bubble structure) in the isotopes of Z = 121 using RMF+BCS approach. In Fig. \ref{fig4}, we have displayed charge density and neutron density of $^{293}$121 and $^{292}$120. Since the nucleus $^{292}$120 is found to be a potential candidate of semi-bubble structure by various theoretical works \cite{Berger2001,Decharge2003,Afanasjev2005,SinghSK2013,Li2016,Schuetrumpf2017} and shows fairly good agreement with our calculations for Z $=$ 120, we extend similar calculations to Z = 121 isotopes to examine the bubble structures. The depletion in charge density is observed in various isotopes of Z $=$ 121 mostly on the neutron deficient side where many nuclei are proton unbound and lie beyond proton drip line. In Fig. \ref{fig4}, we show depletion in charge density for $^{293}$121 which is relatively stable and is the first proton bound nucleus.\par
The depletion fraction (D. F.) $=$ ($\rho_{max}-\rho_{c})/\rho_{max}$ \cite{Grasso2009} computed for Z = 121 isotopes  vs. N is shown in Fig. \ref{fig5} and the effect of neutron number variation (if any) on depletion fraction is investigated. We find that the depletion in the charge density, which is mainly because of the Coulomb repulsion, is maximum on the neutron deficient side. As neutron number increases,  the depletion fraction starts decreasing as the excess number of neutrons balance the Coulomb repulsion and decrease the depletion in charge density which is evident in Fig. \ref{fig5}. This kind of variation in depletion fraction with respect to neutron number is expected to occur for all the superheavy nuclei which needs further investigations on the bubble phenomenon.
\subsection{Decay modes of $^{293-312}${121}}
Investigation of decay properties is one dominant way to probe superheavy nuclei and their stability. $\alpha$-decay and spontaneous fission (SF) have achieved a great success during the last two decades for the identification of new elements \cite{Hofmann2016,Oganessian2009,Bao2015,Wang2015,Niyti2015,Heenen2015,Santhosh2016,Budaca2016,Liu2017,Zhang2017} whereas the competition between SF and $\alpha$-decay plays a crucial role in the detection of superheavy nuclei in the laboratory. However, $\alpha$-decay is found as a very powerful tool to investigate the nuclear structure properties like shell effects, nuclear spins and parities, deformation, rotational properties and fission barrier etc. of superheavy nuclei.\par
Here we present a systematic study of the decay properties of superheavy nuclei and probe the competition between $\alpha$-decay and spontaneous fission (SF) by calculating the $\alpha$-decay half-lives and spontaneous fission half-lives of Z $=$ 121 isotopes with A $=$ 293 - 380 using RMF+BCS and NSM approaches with various formulas given in the Refs. \cite{VSS1966,Brown1992,Sobiczewski2005,Qi2009,Ni2011,Akrawy2017,Xu2008}. In order to compare and validate our results, we also compute Q-values and $log_{10}T_{\alpha}$ for the decay chains of superheavy nuclei $^{294}$117 and $^{293}$117 which have already been synthesized and their $\alpha$-decay chains have been reported \cite{Oganessian2010}.
\begin{figure}
\resizebox{0.5\textwidth}{!}{%
  \includegraphics{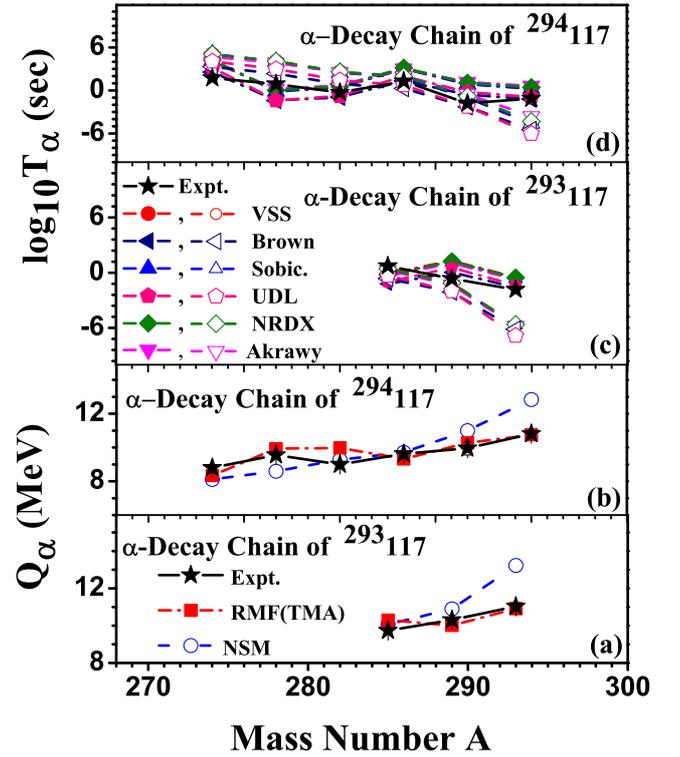}
}
\caption{(Colour online) Calculated Q-values and $\alpha$-decay half-lives for decay chain of $^{294}$117 and $^{293}$117 are compared with experimental values \cite{Oganessian2010}. RMF and NSM data are denoted by filled and opaque symbols, respectively.}
\label{fig6}
\end{figure}
Our calculated values of Q$_{\alpha}$ and log$_{10}T_{\alpha}$ for decay chains of $^{294}$117 and $^{293}$117  using various formulas (see Appendix for details of all the formulas) given by Viola and Seaborg(VSS) \cite{VSS1966}, Brown \cite{Brown1992}, Sobiczewski \cite{Sobiczewski2005}, universal decay law (UDL) introduced by Qi \textit{et al.} \cite{Qi2009}, unified formula for ${\alpha}$-decay and cluster decay \cite{Ni2011}, Royer formula given by Akrawy \textit{et al.} \cite{Akrawy2017}, are shown in Fig. \ref{fig6} and compared  with the experimental values \cite{Oganessian2010}. Both the theories (RMF and NSM) are able to reproduce the experimental data reasonably well in the superheavy region that shows efficacy of our calculations.
\begin{figure}
\resizebox{0.5\textwidth}{!}{%
  \includegraphics{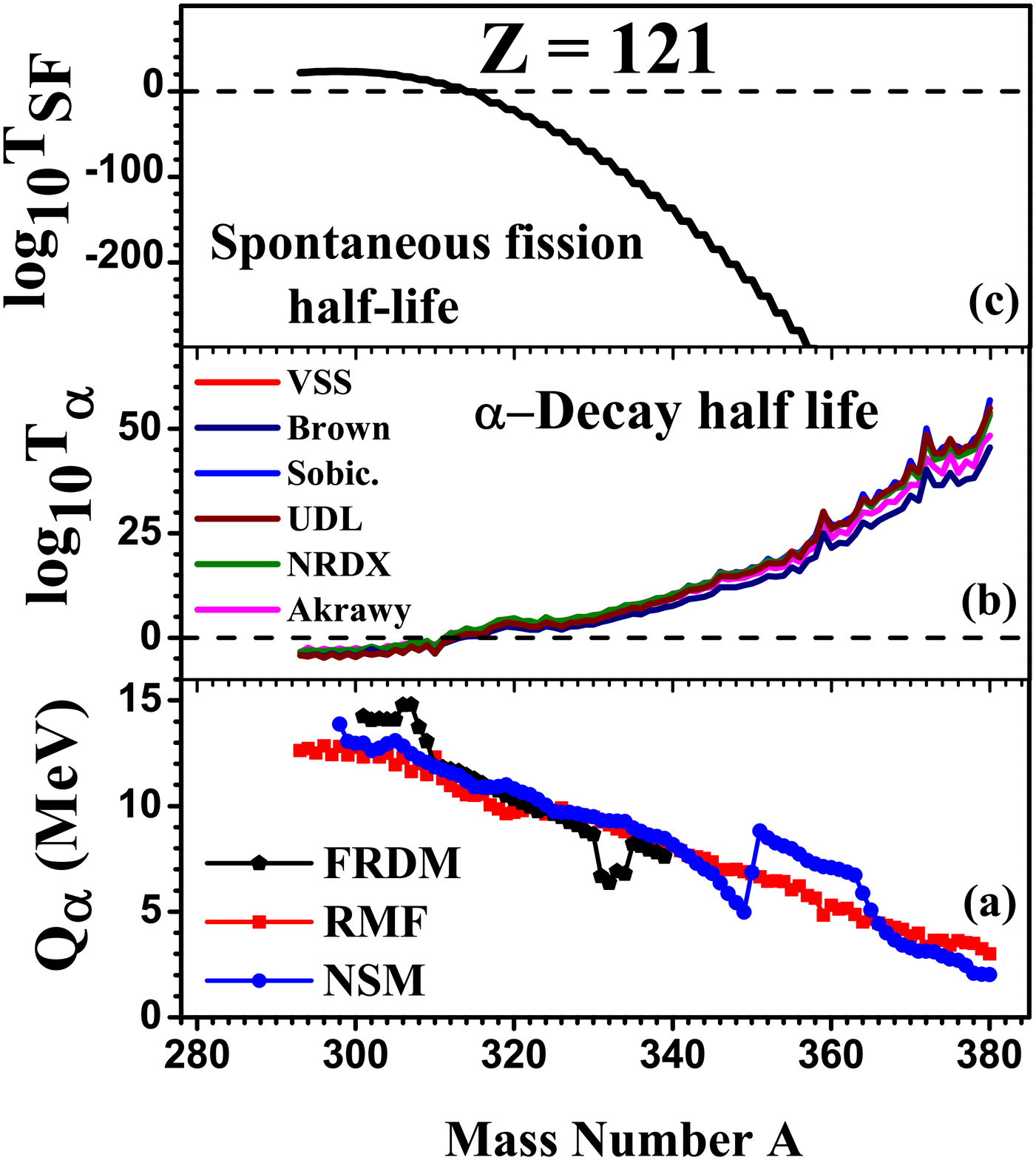}
}
\caption{(Colour online) Calculated Q-values, $\alpha$-decay half-lives (sec) and spontaneous fission half-lives (sec) vs A for Z = 121 isotopes.}
\label{fig7}
\end{figure}
Fig. \ref{fig7}(a) displays our calculated values (using RMF and NSM)  of Q$_{\alpha}$ for Z = 121 isotopic chain (A = 293 - 380) as a function of mass number A along with the results of FRDM \cite{Moller2012} which show reasonable agreement with our work. Fig. \ref{fig7}(b) shows $log_{10}T_{\alpha}$ calculated using various formula \cite{VSS1966,Brown1992,Sobiczewski2005,Qi2009,Ni2011,Akrawy2017}. In Fig. \ref{fig7}(c), we plot the spontaneous fission half-life (T$_{SF}$) calculated using the formula of C. Xu $\textit{et al.}$ \cite{Xu2008} given below and study the competition between  $\alpha$-decay and SF decay modes.

\begin{eqnarray}
T_{SF}(1/2) = exp[2\pi\{C_{0} + C_{1}A + C_{2}Z^2 + C_{3}Z^4 \nonumber\\ +  C_{4}(N-Z)^2 -(0.13323\frac{Z^2}{A^{1/3}} - 11.64)\}]
\end{eqnarray}
The constants are $C_0$ = -195.09227, $C_1$ = 3.10156, $C_2$ = -0.04386, $C_3$ = 1.4030$\times10^{-6}$, and $C_4$ = -0.03199.\par

As is seen from the Figs. \ref{fig7}(b) and \ref{fig7}(c), spontaneous fission half-life (T$_{SF}$) is larger than $\alpha$-decay half-life ($log_{10}T_{\alpha}$) for lower mass A $=$ 293 $-$ 312, which shows that the $\alpha$-decay is found to be favourable decay mode for A $<$ 312 whereas for the isotopes with A $>$ 312, the decay by spontaneous fission is more favourable. This is worthy to mention here that isotopes between A $=$ 293 $-$ 312 have B.E/A $>$ 7 MeV and found at the top of plot of B.E/A curve and therefore are found most stable among full isotopic chain of Z = 121. We have calculated average of B.E/A of the around 80 nuclei which are experimentally known in this region so far and it is quite satisfactory to note that this average is found 7.2 MeV very close to B.E/A for our predicted isotopes of Z = 121 which are expected to decay through ${\alpha}$-decay. \par

Table 1 \& 2 show the calculated values of Q$_{\alpha}$, $\alpha$-decay half-life (T$_{\alpha}$) from various formula \cite{VSS1966,Brown1992,Sobiczewski2005,Qi2009,Ni2011,Akrawy2017}, spontaneous fission half-life (T$_{SF}$) with the formula of C. Xu $\textit{et al.}$ \cite{Xu2008} and the possible decay mode which could be either $\alpha$ decay or SF. In Table 3, we have compared Q$_{\alpha}$, $\alpha$-decay half-life (T$_{\alpha}$) as well as the decay mode with the available experimental data taken from Ref. \cite{Oganessian2015}. It may be noted from these tables that the nuclei between A $=$ 293 $-$ 302 (approximately 10 potential candidates) are found with long $\alpha$-decay chain for which our calculated $\alpha$-decay half-life and predicted decay mode are in excellent agreement with available data from experiments \cite{Oganessian2015}. The decay chains of nuclei with 303 $\leq$ A $\leq$ 312 are terminated by SF after 3$\alpha$/2$\alpha$/1$\alpha$ before reaching the known territory of the nuclear chart. Therefore such nuclei are still far from the reach of experiments as of now. However, nuclei with A = 293 - 302 (N = 172 - 281) are found to have enough potential to be observed or produced experimentally.
\begin{table*}[htbp]
\caption{Predictions on the modes of decay of $^{293-312}${121} superheavy nuclei and their decay products (decay-chain) by comparing the alpha half-lives (sec) and the corresponding SF half-lives (sec). The half-lives are calculated using formula of Viola and Seaborg (VSS) \cite{VSS1966}, B. A. Brown (Brown) \cite{Brown1992}, Sobiczewski $\textit{et al.}$ \cite{Sobiczewski2005}, universal decay law (UDL) \cite{Qi2009}, unified formula (NRDX) \cite{Ni2011}, modified Royer formula given by Akrawy $\textit{et al.}$ \cite{Akrawy2017} and formula of Xu $\textit{et al.}$ for spontaneous fission \cite{Xu2008}.}
\centering
\resizebox{1.0\textwidth}{!}{%
%\textbf{
{\begin{tabular}{cccccccccc}
 \hline
 \hline\multicolumn{1}{c}{Nuclei}&
 \multicolumn{1}{c}{Q$_{\alpha}${RMF}}&
 \multicolumn{6}{c}{T$_{\alpha}${(1/2)}}&
 \multicolumn{1}{c}{T$_{SF}${(1/2)}}&
 \multicolumn{1}{c}{Decay Mode}\\
\cline{3-8}
 \multicolumn{1}{c}{}&
 \multicolumn{1}{c}{(MeV)}&
 \multicolumn{1}{c}{VSS}&
  \multicolumn{1}{c}{Brown}&
   \multicolumn{1}{c}{Sobiczewski $\textit{et al.}$}&
    \multicolumn{1}{c}{UDL}&
     \multicolumn{1}{c}{NRDX}&
      \multicolumn{1}{c}{Akrawy $\textit{et al.}$}&
       \multicolumn{1}{c}{C. Xu $\textit{et al.}$}&
        \multicolumn{1}{c}{RMF}\\
 \hline
$^{293}${121}& 12.63&	3.74$\times$10$^{-04}$&	7.20$\times$10$^{-05}$&	3.61$\times$10$^{-04}$&	5.68$\times$10$^{-05}$&	 4.66$\times$10$^{-04}$&	 3.30$\times$10$^{-04}$&	 1.22$\times$10$^{+22 }$&$\alpha1$\\
$^{289}${119}& 11.72&	1.17$\times$10$^{-02}$&	1.43$\times$10$^{-03}$&	9.87$\times$10$^{-03}$&	2.00$\times$10$^{-03}$&	 1.26$\times$10$^{-02}$&	 9.28$\times$10$^{-03}$&	 5.77$\times$10$^{+14 }$&$\alpha2$\\
$^{285}${Ts}& 13.02&	5.16$\times$10$^{-06}$&	1.55$\times$10$^{-06}$&	4.77$\times$10$^{-06}$&	5.19$\times$10$^{-07}$&	 5.91$\times$10$^{-06}$&	 4.64$\times$10$^{-06}$&	 7.23$\times$10$^{+08}$ &$\alpha3$\\
$^{281}${Mc}& 12.83&	3.81$\times$10$^{-06}$&	1.14$\times$10$^{-06}$&	3.33$\times$10$^{-06}$&	3.51$\times$10$^{-07}$&	 4.11$\times$10$^{-06}$&	 3.36$\times$10$^{-06}$&	 1.97$\times$10$^{+04}$ &$\alpha4$ \\
$^{277}${Nh}& 11.46&	8.99$\times$10$^{-04}$&	1.49$\times$10$^{-04}$&	6.72$\times$10$^{-04}$&	1.08$\times$10$^{-04}$&	 8.35$\times$10$^{-04}$&	 7.05$\times$10$^{-04}$&	 9.84$\times$10$^{+00}$ &$\alpha5$\\
$^{273}${Rg}& 11.47&	2.24$\times$10$^{-04}$&	4.30$\times$10$^{-05}$&	1.62$\times$10$^{-04}$&	2.34$\times$10$^{-05}$&	 2.02$\times$10$^{-04}$&	 1.78$\times$10$^{-04}$&	 9.92$\times$10$^{-02}$ &$\alpha6$\\
$^{269}${Mt}& 11.25&	1.88$\times$10$^{-04}$&	3.72$\times$10$^{-05}$&	1.29$\times$10$^{-04}$&	1.85$\times$10$^{-05}$&	 1.63$\times$10$^{-04}$&	 1.49$\times$10$^{-04}$&	 3.13$\times$10$^{-02}$ &$\alpha7$\\
\\
$^{294}${121}&12.73&	5.01$\times$10$^{-04}$&	4.67$\times$10$^{-05}$&	5.07$\times$10$^{-04}$&	3.24$\times$10$^{-05}$&	 1.02$\times$10$^{-03}$&	 4.60$\times$10$^{-03}$&	 6.14$\times$10$^{+22}$& $\alpha1$ \\
$^{290}${119}&11.69&	3.10$\times$10$^{-02}$&	1.69$\times$10$^{-03}$&	3.00$\times$10$^{-02}$&	2.36$\times$10$^{-03}$&	 5.39$\times$10$^{-02}$&	 1.55$\times$10$^{-01}$&	 2.74$\times$10$^{+15}$& $\alpha2$\\
$^{286}${Ts}&12.84&	2.66$\times$10$^{-05}$&	3.30$\times$10$^{-06}$&	2.42$\times$10$^{-05}$&	1.23$\times$10$^{-06}$&	 4.85$\times$10$^{-05}$&	 2.70$\times$10$^{-04}$&	 3.24$\times$10$^{+09}$&$\alpha3$\\
$^{282}${Mc}&12.88&	6.61$\times$10$^{-06}$&	9.23$\times$10$^{-07}$&	5.70$\times$10$^{-06}$&	2.63$\times$10$^{-07}$&	 1.15$\times$10$^{-05}$&	 7.06$\times$10$^{-05}$&	 8.32$\times$10$^{+04}$&$\alpha4$ \\
$^{278}${Nh}&11.34&	3.93$\times$10$^{-03}$&	2.76$\times$10$^{-04}$&	3.32$\times$10$^{-03}$&	2.16$\times$10$^{-04}$&	 5.82$\times$10$^{-03}$&	 1.81$\times$10$^{-02}$&	 3.82$\times$10$^{+01}$&$\alpha5$\\
$^{274}${Rg}&11.22&	1.95$\times$10$^{-03}$&	1.50$\times$10$^{-04}$&	1.58$\times$10$^{-03}$&	9.74$\times$10$^{-05}$&	 2.79$\times$10$^{-03}$&	 8.87$\times$10$^{-03}$&	 2.86$\times$10$^{-01}$&$\alpha6$\\
$^{270}${Mt}&10.06&	3.91$\times$10$^{-01}$&	1.93$\times$10$^{-02}$&	3.24$\times$10$^{-01}$&	2.58$\times$10$^{-02}$&	 4.94$\times$10$^{-01}$&	 8.92$\times$10$^{-01}$&	 4.70$\times$10$^{-02}$&SF\\
\\
$^{295}${121}& 12.51&	6.77$\times$10$^{-04}$&	1.21$\times$10$^{-04}$&	6.45$\times$10$^{-04}$&	9.88$\times$10$^{-05}$&	 8.34$\times$10$^{-04}$&	 5.73$\times$10$^{-04}$&	 1.10$\times$10$^{+23}$ &$\alpha1$\\
$^{291}${119}& 11.55&	2.96$\times$10$^{-02}$&	3.25$\times$10$^{-03}$&	2.47$\times$10$^{-02}$&	5.00$\times$10$^{-03}$&	 3.15$\times$10$^{-02}$&	 2.24$\times$10$^{-02}$&	 4.90$\times$10$^{+15}$ &$\alpha2$\\
$^{287}${Ts}& 11.86&	1.56$\times$10$^{-03}$&	2.43$\times$10$^{-04}$&	1.29$\times$10$^{-03}$&	2.05$\times$10$^{-04}$&	 1.62$\times$10$^{-03}$&	 1.22$\times$10$^{-03}$&	 5.76$\times$10$^{+09}$ &$\alpha3$\\
$^{283}${Mc}& 13.07&	1.27$\times$10$^{-06}$&	4.28$\times$10$^{-07}$&	1.13$\times$10$^{-06}$&	1.01$\times$10$^{-07}$&	 1.39$\times$10$^{-06}$&	 1.10$\times$10$^{-06}$&	 1.47$\times$10$^{+05}$ &$\alpha4$\\
$^{279}${Nh}& 11.23&	3.11$\times$10$^{-03}$&	4.55$\times$10$^{-04}$&	2.28$\times$10$^{-03}$&	3.75$\times$10$^{-04}$&	 2.84$\times$10$^{-03}$&	 2.30$\times$10$^{-03}$&	 6.66$\times$10$^{+01}$ &$\alpha5$\\
$^{275}${Rg}& 11.01&	2.74$\times$10$^{-03}$&	4.15$\times$10$^{-04}$&	1.91$\times$10$^{-03}$&	3.10$\times$10$^{-04}$&	 2.38$\times$10$^{-03}$&	 2.01$\times$10$^{-03}$&	 4.72$\times$10$^{-01}$ &$\alpha6$\\
$^{271}${Mt}& 11.05&	5.43$\times$10$^{-04}$&	9.79$\times$10$^{-05}$&	3.68$\times$10$^{-04}$&	5.30$\times$10$^{-05}$&	 4.65$\times$10$^{-04}$&	 4.06$\times$10$^{-04}$&	 6.26$\times$10$^{-02}$ &$\alpha7$\\
\\
$^{296}${121}&12.85&	2.77$\times$10$^{-04}$&	2.79$\times$10$^{-05}$&	2.81$\times$10$^{-04}$&	1.61$\times$10$^{-05}$&	 5.69$\times$10$^{-04}$&	 2.58$\times$10$^{-03}$&	 1.51$\times$10$^{+23}$ &$\alpha1$\\
$^{292}${119}&11.37&	1.81$\times$10$^{-01}$&	7.98$\times$10$^{-03}$&	1.76$\times$10$^{-01}$&	1.42$\times$10$^{-02}$&	 3.05$\times$10$^{-01}$&	 6.99$\times$10$^{-01}$&	 6.46$\times$10$^{+15}$ &$\alpha2$\\
$^{288}${Ts}&11.64&	1.06$\times$10$^{-02}$&	6.60$\times$10$^{-04}$&	9.80$\times$10$^{-03}$&	6.55$\times$10$^{-04}$&	 1.75$\times$10$^{-02}$&	 5.12$\times$10$^{-02}$&	 7.31$\times$10$^{+09}$ &$\alpha3$\\
$^{284}${Mc}&13.11&	2.39$\times$10$^{-06}$&	3.72$\times$10$^{-07}$&	2.06$\times$10$^{-06}$&	8.28$\times$10$^{-08}$&	 4.25$\times$10$^{-06}$&	 2.73$\times$10$^{-05}$&	 1.79$\times$10$^{+05}$ &$\alpha4$\\
$^{280}${Nh}&10.56&	3.20$\times$10$^{-01}$&	1.44$\times$10$^{-02}$&	2.80$\times$10$^{-01}$&	2.14$\times$10$^{-02}$&	 4.42$\times$10$^{-01}$&	 8.51$\times$10$^{-01}$&	 7.82$\times$10$^{+01}$ &$\alpha5$\\
$^{276}${Rg}&11.26&	1.52$\times$10$^{-03}$&	1.20$\times$10$^{-04}$&	1.23$\times$10$^{-03}$&	6.95$\times$10$^{-05}$&	 2.19$\times$10$^{-03}$&	 6.83$\times$10$^{-03}$&	 5.15$\times$10$^{-01}$ &$\alpha6$\\
$^{272}${Mt}& 9.88&1.16$\times$10$^{+00}$	&5.19$\times$10$^{-02}$	&9.74$\times$10$^{-01}$	&7.58$\times$10$^{-02}$	 &1.44$\times$10$^{+00}$	 &2.24$\times$10$^{+00}$	 &5.45$\times$10$^{-02}$  &SF\\
\\
$^{297}${121}& 12.44&	9.72$\times$10$^{-04}$&	1.65$\times$10$^{-04}$&	9.20$\times$10$^{-04}$&	1.35$\times$10$^{-04}$&	 1.19$\times$10$^{-03}$&	 7.89$\times$10$^{-04}$&	 1.92$\times$10$^{+23}$&$\alpha1$ \\
$^{293}${119}& 11.39&	7.35$\times$10$^{-02}$&	7.22$\times$10$^{-03}$&	6.01$\times$10$^{-02}$&	1.22$\times$10$^{-02}$&	 7.71$\times$10$^{-02}$&	 5.28$\times$10$^{-02}$&	 8.01$\times$10$^{+15}$&$\alpha2$\\
$^{289}${Ts}& 10.72&	9.71$\times$10$^{-01}$&	7.18$\times$10$^{-02}$&	7.18$\times$10$^{-01}$&	1.74$\times$10$^{-01}$&	 9.06$\times$10$^{-01}$&	 6.44$\times$10$^{-01}$&	 8.86$\times$10$^{+09}$&$\alpha3$\\
$^{285}${Mc}& 12.77&	4.88$\times$10$^{-06}$&	1.42$\times$10$^{-06}$&	4.25$\times$10$^{-06}$&	3.93$\times$10$^{-07}$&	 5.25$\times$10$^{-06}$&	 3.97$\times$10$^{-06}$&	 2.12$\times$10$^{+05}$&$\alpha4$\\
$^{281}${Nh}& 11.19&	3.97$\times$10$^{-03}$&	5.65$\times$10$^{-04}$&	2.89$\times$10$^{-03}$&	4.50$\times$10$^{-04}$&	 3.60$\times$10$^{-03}$&	 2.80$\times$10$^{-03}$&	 8.97$\times$10$^{+01}$&$\alpha5$\\
$^{277}${Rg}& 10.71&	1.47$\times$10$^{-02}$&	1.90$\times$10$^{-03}$&	1.00$\times$10$^{-02}$&	1.71$\times$10$^{-03}$&	 1.25$\times$10$^{-02}$&	 1.00$\times$10$^{-02}$&	 5.58$\times$10$^{-01}$&$\alpha6$\\
$^{273}${Mt}& 10.92&	1.16$\times$10$^{-03}$&	1.96$\times$10$^{-04}$&	7.79$\times$10$^{-04}$&	1.10$\times$10$^{-04}$&	 9.84$\times$10$^{-04}$&	 8.18$\times$10$^{-04}$&	 4.64$\times$10$^{-02}$&$\alpha7$\\
\\
$^{298}${121}&12.81&	3.36$\times$10$^{-04}$&	3.30$\times$10$^{-05}$&	3.41$\times$10$^{-04}$&	1.83$\times$10$^{-05}$&	 6.89$\times$10$^{-04}$&	 2.91$\times$10$^{-03}$&	 1.29$\times$10$^{+23}$&$\alpha1$\\
$^{294}${119}&11.29&	2.78$\times$10$^{-01}$&	1.16$\times$10$^{-02}$&	2.71$\times$10$^{-01}$&	2.08$\times$10$^{-02}$&	 4.66$\times$10$^{-01}$&	 9.74$\times$10$^{-01}$&	 5.27$\times$10$^{+15}$&$\alpha2$\\
$^{290}${Ts}&10.82&	1.14$\times$10$^{+00}$&	4.11$\times$10$^{-02}$&	1.07$\times$10$^{+00}$&	8.62$\times$10$^{-02}$&	 1.73$\times$10$^{+00}$&	 3.04$\times$10$^{+00}$&	 5.75$\times$10$^{+09}$&$\alpha3$\\
$^{286}${Mc}&11.87&	8.90$\times$10$^{-04}$&	7.29$\times$10$^{-05}$&	7.77$\times$10$^{-04}$&	4.10$\times$10$^{-05}$&	 1.44$\times$10$^{-03}$&	 4.93$\times$10$^{-03}$&	 1.35$\times$10$^{+05}$&$\alpha4$\\
$^{282}${Nh}&11.27&	5.65$\times$10$^{-03}$&	3.83$\times$10$^{-04}$&	4.78$\times$10$^{-03}$&	2.74$\times$10$^{-04}$&	 8.35$\times$10$^{-03}$&	 2.30$\times$10$^{-02}$&	 5.66$\times$10$^{+01}$&$\alpha5$\\
$^{278}${Rg}&10.49&	1.16$\times$10$^{-01}$&	6.04$\times$10$^{-03}$&	9.75$\times$10$^{-02}$&	6.42$\times$10$^{-03}$&	 1.56$\times$10$^{-01}$&	 3.06$\times$10$^{-01}$&	 3.47$\times$10$^{-01}$&$\alpha6$\\
$^{274}${Mt}&9.85	&1.47$\times$10$^{+00}$	&6.45$\times$10$^{-02}$	&1.24$\times$10$^{+00}$	&9.05$\times$10$^{-02}$	 &1.82$\times$10$^{+00}$	 &2.68$\times$10$^{+00}$	 &2.80$\times$10$^{-02}$ &$\alpha7$/SF\\
\\
$^{299}${121}&12.42&	1.04$\times$10$^{-03}$&	1.76$\times$10$^{-04}$&	9.86$\times$10$^{-04}$&	1.35$\times$10$^{-04}$&	 1.28$\times$10$^{-03}$&	 8.14$\times$10$^{-04}$&	 6.48$\times$10$^{+22}$&$\alpha1$\\
$^{295}${119}&11.17&	2.58$\times$10$^{-01}$&	2.17$\times$10$^{-02}$&	2.06$\times$10$^{-01}$&	4.27$\times$10$^{-02}$&	 2.65$\times$10$^{-01}$&	 1.73$\times$10$^{-01}$&	 2.53$\times$10$^{+15}$&$\alpha2$\\
$^{291}${Ts}&10.72&	9.82$\times$10$^{-01}$&	7.26$\times$10$^{-02}$&	7.26$\times$10$^{-01}$&	1.64$\times$10$^{-01}$&	 9.19$\times$10$^{-01}$&	 6.25$\times$10$^{-01}$&	 2.63$\times$10$^{+09}$&$\alpha3$\\
$^{287}${Mc}&11.21&	1.37$\times$10$^{-02}$&	1.68$\times$10$^{-03}$&	1.03$\times$10$^{-02}$&	1.65$\times$10$^{-03}$&	 1.29$\times$10$^{-02}$&	 9.20$\times$10$^{-03}$&	 5.91$\times$10$^{+04}$&$\alpha4$\\
$^{283}${Nh}&11.00&	1.14$\times$10$^{-02}$&	1.46$\times$10$^{-03}$&	8.18$\times$10$^{-03}$&	1.28$\times$10$^{-03}$&	 1.02$\times$10$^{-02}$&	 7.54$\times$10$^{-03}$&	 2.34$\times$10$^{+01}$&$\alpha5$\\
$^{279}${Rg}&10.52&	4.53$\times$10$^{-02}$&	5.25$\times$10$^{-03}$&	3.03$\times$10$^{-02}$&	5.25$\times$10$^{-03}$&	 3.78$\times$10$^{-02}$&	 2.89$\times$10$^{-02}$&	 1.35$\times$10$^{-01}$&$\alpha6$\\
$^{275}${Mt}&10.90&	1.29$\times$10$^{-03}$&	2.16$\times$10$^{-04}$&	8.68$\times$10$^{-04}$&	1.15$\times$10$^{-04}$&	 1.10$\times$10$^{-03}$&	 8.67$\times$10$^{-04}$&	 9.52$\times$10$^{-03}$&$\alpha7$\\
\\
$^{300}${121}&12.76&	4.24$\times$10$^{-04}$&	4.04$\times$10$^{-05}$&	4.30$\times$10$^{-04}$&	2.18$\times$10$^{-05}$&	 8.67$\times$10$^{-04}$&	 3.42$\times$10$^{-03}$&	 3.45$\times$10$^{+22}$&$\alpha1$\\
$^{296}${119}&11.11&	8.19$\times$10$^{-01}$&	3.00$\times$10$^{-02}$&	8.03$\times$10$^{-01}$&	6.08$\times$10$^{-02}$&	 1.35$\times$10$^{+00}$&	 2.42$\times$10$^{+00}$&	 1.34$\times$10$^{+15}$&$\alpha2$\\
$^{292}${Ts}&10.93&	5.96$\times$10$^{-01}$&	2.33$\times$10$^{-02}$&	5.62$\times$10$^{-01}$&	4.04$\times$10$^{-02}$&	 9.21$\times$10$^{-01}$&	 1.65$\times$10$^{+00}$&	 1.39$\times$10$^{+09}$&$\alpha3$\\
$^{288}${Mc}&10.13&	2.06$\times$10$^{+01}$&	5.66$\times$10$^{-01}$&	1.94$\times$10$^{+01}$&	1.62$\times$10$^{+00}$&	 2.82$\times$10$^{+01}$&	 3.44$\times$10$^{+01}$&	 3.11$\times$10$^{+04}$&$\alpha4$\\
$^{284}${Nh}&10.82&	6.82$\times$10$^{-02}$&	3.59$\times$10$^{-03}$&	5.88$\times$10$^{-02}$&	3.57$\times$10$^{-03}$&	 9.70$\times$10$^{-02}$&	 2.01$\times$10$^{-01}$&	 1.23$\times$10$^{+01}$&$\alpha5$\\
$^{280}${Rg}&10.43&	1.67$\times$10$^{-01}$&	8.41$\times$10$^{-03}$&	1.41$\times$10$^{-01}$&	8.79$\times$10$^{-03}$&	 2.24$\times$10$^{-01}$&	 4.10$\times$10$^{-01}$&	 7.08$\times$10$^{-02}$&SF\\
$^{276}${Mt}&9.99&  6.02$\times$10$^{-01}$	&2.86$\times$10$^{-02}$	&5.02$\times$10$^{-01}$	&3.27$\times$10$^{-02}$	 &7.60$\times$10$^{-01}$	 &1.18$\times$10$^{+00}$	 &4.96$\times$10$^{-03}$ &SF\\
\\
\hline
\hline
\end{tabular}}}
\end{table*}

\begin{table*}[htbp]
\caption{Table 1 continues}
\centering
\resizebox{1.0\textwidth}{!}{%
%\textbf{
{\begin{tabular}{llllllllll}
 \hline
 \hline\multicolumn{1}{c}{Nuclei}&
 \multicolumn{1}{l}{Q$_{\alpha}${RMF}}&
 \multicolumn{6}{c}{T$_{\alpha}${(1/2)}}&
 \multicolumn{1}{l}{T$_{SF}${(1/2)}}&
 \multicolumn{1}{l}{Decay Mode}\\
\cline{3-8}
 \multicolumn{1}{l}{}&
 \multicolumn{1}{l}{(MeV)}&
 \multicolumn{1}{l}{VSS}&
  \multicolumn{1}{l}{Brown}&
   \multicolumn{1}{l}{Sobiczewski $\textit{et al.}$}&
    \multicolumn{1}{l}{UDL}&
     \multicolumn{1}{l}{NRDX}&
      \multicolumn{1}{l}{Akrawy $\textit{et al.}$}&
       \multicolumn{1}{l}{C. Xu $\textit{et al.}$}&
        \multicolumn{1}{l}{RMF}\\
 \hline
$^{301}${121}& 12.33&	1.69$\times$10$^{-03}$&	2.68$\times$10$^{-04}$&	1.58$\times$10$^{-03}$&	2.09$\times$10$^{-04}$&	 2.06$\times$10$^{-03}$&	 1.25$\times$10$^{-03}$&	 4.21$\times$10$^{+21}$&$\alpha1$\\
$^{297}${119}& 10.86&	1.64$\times$10$^{+00}$&	1.10$\times$10$^{-01}$&	1.27$\times$10$^{+00}$&	2.83$\times$10$^{-01}$&	 1.64$\times$10$^{+00}$&	 1.02$\times$10$^{+00}$&	 1.55$\times$10$^{+14}$&$\alpha2$\\
$^{293}${Ts}& 10.91&	3.02$\times$10$^{-01}$&	2.56$\times$10$^{-02}$&	2.28$\times$10$^{-01}$&	4.37$\times$10$^{-02}$&	 2.89$\times$10$^{-01}$&	 1.87$\times$10$^{-01}$&	 1.51$\times$10$^{+08}$&$\alpha3$\\
$^{289}${Mc}& 10.01&	2.00$\times$10$^{+01}$&	1.11$\times$10$^{+00}$&	1.34$\times$10$^{+01}$&	3.49$\times$10$^{+00}$&	 1.67$\times$10$^{+01}$&	 1.12$\times$10$^{+01}$&	 3.18$\times$10$^{+03}$&$\alpha4$\\
$^{285}${Nh}& 10.30&	7.44$\times$10$^{-01}$&	6.21$\times$10$^{-02}$&	5.00$\times$10$^{-01}$&	1.00$\times$10$^{-01}$&	 6.23$\times$10$^{-01}$&	 4.34$\times$10$^{-01}$& 1.19$\times$10$^{+00}$&$\alpha5$/SF\\
$^{281}${Rg}& 10.37&	1.15$\times$10$^{-01}$&	1.22$\times$10$^{-02}$&	7.58$\times$10$^{-02}$&	1.31$\times$10$^{-02}$&	 9.46$\times$10$^{-02}$&	 6.83$\times$10$^{-02}$&	 6.41$\times$10$^{-03}$&SF\\
$^{277}${Mt}& 10.85&	1.71$\times$10$^{-03}$&	2.79$\times$10$^{-04}$&	1.14$\times$10$^{-03}$&	1.43$\times$10$^{-04}$&	 1.45$\times$10$^{-03}$&	 1.08$\times$10$^{-03}$&	 4.14$\times$10$^{-04}$&SF\\
\\
$^{302}${121}& 12.59&9.83$\times$10$^{-04}$&	8.41$\times$10$^{-05}$&	9.95$\times$10$^{-04}$&	4.93$\times$10$^{-05}$&	 1.98$\times$10$^{-03}$&	 6.90$\times$10$^{-03}$&	 2.13$\times$10$^{+21}$&$\alpha1$\\
$^{298}${119}& 10.58&2.11$\times$10$^{+01}$&	5.20$\times$10$^{-01}$&	2.11$\times$10$^{+01}$&	1.77$\times$10$^{+00}$&	 3.29$\times$10$^{+01}$&	 4.11$\times$10$^{+01}$&	 7.83$\times$10$^{+13}$&$\alpha2$\\
$^{294}${Ts}& 10.75&1.77$\times$10$^{+00}$&	6.08$\times$10$^{-02}$&	1.68$\times$10$^{+00}$&	1.19$\times$10$^{-01}$&	 2.68$\times$10$^{+00}$&	 4.16$\times$10$^{+00}$&	 7.63$\times$10$^{+07}$&$\alpha3$\\
$^{290}${Mc}& 10.27&8.30$\times$10$^{+00}$&	2.51$\times$10$^{-01}$&	7.72$\times$10$^{+00}$&	5.75$\times$10$^{-01}$&	 1.15$\times$10$^{+01}$&	 1.49$\times$10$^{+01}$&	 1.61$\times$10$^{+03}$&$\alpha4$\\
$^{286}${Nh}& 9.33	&1.10$\times$10$^{+03}$&	2.16$\times$10$^{+01}$&	1.07$\times$10$^{+03}$&	9.72$\times$10$^{+01}$&	 1.34$\times$10$^{+03}$&	 1.04$\times$10$^{+03}$&	 5.99$\times$10$^{-01}$&SF\\
$^{282}${Rg}& 9.98	&2.87$\times$10$^{+00}$&	1.10$\times$10$^{-01}$&	2.5$\times$10$^{+00}$&	1.67$\times$10$^{-01}$&	 3.67$\times$10$^{+00}$&	 4.94$\times$10$^{+00}$&	 3.23$\times$10$^{-03}$&SF\\
$^{278}${Mt}& 9.93	&8.41$\times$10$^{-01}$&	3.88$\times$10$^{-02}$&	7.06$\times$10$^{-01}$&	4.33$\times$10$^{-02}$&	 1.06$\times$10$^{+00}$&	 1.56$\times$10$^{+00}$&	 2.09$\times$10$^{-04}$&SF\\
\\
$^{303}${121}& 12.33&	1.69$\times$10$^{-03}$&	2.68$\times$10$^{-04}$&	1.58$\times$10$^{-03}$&	1.95$\times$10$^{-04}$&	 2.06$\times$10$^{-03}$&	 1.19$\times$10$^{-03}$&	 5.29$\times$10$^{+19}$&$\alpha1$\\
$^{299}${119}& 10.35&	4.00$\times$10$^{+01}$&	1.82$\times$10$^{+00}$&	2.92$\times$10$^{+01}$&	7.75$\times$10$^{+00}$&	 3.78$\times$10$^{+01}$&	 2.21$\times$10$^{+01}$&	 1.83$\times$10$^{+12}$&$\alpha2$ \\
$^{295}${Ts}& 10.79&	6.23$\times$10$^{-01}$&	4.85$\times$10$^{-02}$&	4.64$\times$10$^{-01}$&	8.75$\times$10$^{-02}$&	 5.89$\times$10$^{-01}$&	 3.63$\times$10$^{-01}$&	 1.68$\times$10$^{+06}$&$\alpha3$\\
$^{291}${Mc}& 9.93	&3.48$\times$10$^{+01}$	&1.82$\times$10$^{+00}$	&2.31$\times$10$^{+01}$	&5.84$\times$10$^{+00}$	 &2.89$\times$10$^{+01}$	 &1.83$\times$10$^{+01}$	 &3.32$\times$10$^{+01}$ &SF\\
$^{287}${Nh}& 9.13	&2.14$\times$10$^{+03}$	&7.94$\times$10$^{+01}$	&1.29$\times$10$^{+03}$	&4.37$\times$10$^{+02}$	 &1.58$\times$10$^{+03}$	 &1.03$\times$10$^{+03}$	 &1.16$\times$10$^{-02}$ &SF\\
$^{283}${Rg}& 10.02&	1.00$\times$10$^{+00}$&	8.67$\times$10$^{-02}$&	6.42$\times$10$^{-01}$&	1.21$\times$10$^{-01}$&	 8.00$\times$10$^{-01}$&	 5.44$\times$10$^{-01}$&	 5.88$\times$10$^{-05}$&SF\\
$^{279}${Mt}& 11.07&	4.98$\times$10$^{-04}$&	9.05$\times$10$^{-05}$&	3.38$\times$10$^{-04}$&	3.60$\times$10$^{-05}$&	 4.30$\times$10$^{-04}$&	 3.02$\times$10$^{-04}$&	 3.56$\times$10$^{-06}$&SF\\
\\
$^{304}${121}& 12.49&1.64$\times$10$^{-03}$	&1.31$\times$10$^{-04}$&	1.66$\times$10$^{-03}$&	7.86$\times$10$^{-05}$&	 3.27$\times$10$^{-03}$&	 1.04$\times$10$^{-02}$&	 2.65$\times$10$^{+19 }$&$\alpha1$\\
$^{300}${119}& 10.12&4.26$\times$10$^{+02}$	&7.27$\times$10$^{+00}$&	4.36$\times$10$^{+02}$&	3.97$\times$10$^{+01}$&	 6.32$\times$10$^{+02}$&	 5.65$\times$10$^{+02}$&	 9.16$\times$10$^{+11 }$&$\alpha2$\\
$^{296}${Ts}& 10.49&8.50$\times$10$^{+00}$	&2.44$\times$10$^{-01}$&	8.16$\times$10$^{+00}$&	5.84$\times$10$^{-01}$&	 1.26$\times$10$^{+01}$&	 1.62$\times$10$^{+01}$&	 8.39$\times$10$^{+05 }$&$\alpha3$\\
$^{292}${Mc}& 10.22&1.17$\times$10$^{E+01}$	&3.42$\times$10$^{-01}$&	1.10$\times$10$^{+01}$&	7.72$\times$10$^{-01}$&	 1.63$\times$10$^{+01}$&	 1.97$\times$10$^{+01}$&	 1.66$\times$10$^{+01}$ &SF\\
$^{288}${Nh}& 8.80	&6.15$\times$10$^{+04}$	&7.98$\times$10$^{+02}$&	6.34$\times$10$^{+04}$&	6.43$\times$10$^{+03}$&	 7.03$\times$10$^{+04}$&	 3.55$\times$10$^{+04}$&	 5.81$\times$10$^{-03}$ &SF\\
$^{284}${Rg}& 9.24	&4.18$\times$10$^{+02}$	&9.98$\times$10$^{+00}$&	3.91$\times$10$^{+02}$&	3.06$\times$10$^{+01}$&	 4.96$\times$10$^{+02}$&	 3.99$\times$10$^{+02}$&	 2.95$\times$10$^{-05}$ &SF\\
$^{280}${Mt}& 9.59	&7.81$\times$10$^{+00}$	&2.97$\times$10$^{-01}$&	6.77$\times$10$^{+00}$&	4.29$\times$10$^{-01}$&	 9.52$\times$10$^{+00}$&	 1.10$\times$10$^{+01}$&	 1.78$\times$10$^{-06}$ &SF\\
\\
$^{304}${121}& 12.49&1.64$\times$10$^{-03}$	&1.31$\times$10$^{-04}$&	1.66$\times$10$^{-03}$&	7.86$\times$10$^{-05}$&	 3.27$\times$10$^{-03}$&	 1.04$\times$10$^{-02}$&	 2.65$\times$10$^{+19 }$&$\alpha1$ \\
$^{300}${119}& 10.12&4.26$\times$10$^{+02}$	&7.27$\times$10$^{+00}$&	4.36$\times$10$^{+02}$&	3.97$\times$10$^{+01}$&	 6.32$\times$10$^{+02}$&	 5.65$\times$10$^{+02}$&	 9.16$\times$10$^{+11 }$&$\alpha2$ \\
$^{296}${Ts}& 10.49&8.50$\times$10$^{+00}$	&2.44$\times$10$^{-01}$&	8.16$\times$10$^{+00}$&	5.84$\times$10$^{-01}$&	 1.26$\times$10$^{+01}$&	 1.62$\times$10$^{+01}$&	 8.39$\times$10$^{+05 }$&$\alpha3$\\
$^{292}${Mc}& 10.22&1.17$\times$10$^{E+01}$	&3.42$\times$10$^{-01}$&	1.10$\times$10$^{+01}$&	7.72$\times$10$^{-01}$&	 1.63$\times$10$^{+01}$&	 1.97$\times$10$^{+01}$&	 1.66$\times$10$^{+01}$ &SF\\
$^{288}${Nh}& 8.80	&6.15$\times$10$^{+04}$	&7.98$\times$10$^{+02}$&	6.34$\times$10$^{+04}$&	6.43$\times$10$^{+03}$&	 7.03$\times$10$^{+04}$&	 3.55$\times$10$^{+04}$&	 5.81$\times$10$^{-03}$ &SF \\
$^{284}${Rg}& 9.24	&4.18$\times$10$^{+02}$	&9.98$\times$10$^{+00}$&	3.91$\times$10$^{+02}$&	3.06$\times$10$^{+01}$&	 4.96$\times$10$^{+02}$&	 3.99$\times$10$^{+02}$&	 2.95$\times$10$^{-05}$ &  SF  \\
$^{280}${Mt}& 9.59	&7.81$\times$10$^{+00}$	&2.97$\times$10$^{-01}$&	6.77$\times$10$^{+00}$&	4.29$\times$10$^{-01}$&	 9.52$\times$10$^{+00}$&	 1.10$\times$10$^{+01}$&	 1.78$\times$10$^{-06}$ &  SF      \\
\\
$^{306}${121}&12.28&4.81$\times$10$^{-03	}$&3.35$\times$10$^{-04}$	&4.87$\times$10$^{-03}$	&2.29$\times$10$^{-04}$	 &9.44$\times$10$^{-03}$	 &2.62$\times$10$^{-02}$	 &6.43$\times$10$^{+16}$&$\alpha1$\\
$^{302}${119}&9.89	&2.08$\times$10$^{+03	}$&2.92$\times$10$^{+01}$	&2.16$\times$10$^{+03}$	&1.98$\times$10$^{+02}$	 &3.00$\times$10$^{+03}$	 &2.23$\times$10$^{+03}$	 &2.09$\times$10$^{+09}$& $\alpha2$\\
$^{298}${Ts}&9.91	&3.97$\times$10$^{+02}$	&7.31$\times$10$^{+00}$	&3.95$\times$10$^{+02}$	&3.21$\times$10$^{+01}$	 &5.52$\times$10$^{+02}$	 &4.72$\times$10$^{+02}$	 &1.80$\times$10$^{+03}$&$\alpha3$/SF\\
$^{294}${Mc}&9.28	&7.84$\times$10$^{+03}$	&1.13$\times$10$^{+02}$	&7.92$\times$10$^{+03}$	&7.11$\times$10$^{+02}$	 &9.77$\times$10$^{+03}$	 &6.07$\times$10$^{+03}$	 &3.35$\times$10$^{-02}$&SF\\
$^{290}${Nh}&9.63	&1.32$\times$10$^{+02}$	&3.21$\times$10$^{+00}$	&1.24$\times$10$^{+02}$	&8.85$\times$10$^{+00}$	 &1.67$\times$10$^{+02}$	 &1.52$\times$10$^{+02}$	 &1.10$\times$10$^{-05}$&SF\\
$^{286}${Rg}&8.81	&1.09$\times$10$^{+04}$	&1.90$\times$10$^{+02}$	&1.08$\times$10$^{+04}$	&9.06$\times$10$^{+02}$	 &1.23$\times$10$^{+04}$	 &7.04$\times$10$^{+03}$	 &5.23$\times$10$^{-08}$&SF\\
$^{282}${Mt}&8.66	&6.38$\times$10$^{+03}$	&1.35$\times$10$^{+02}$	&6.24$\times$10$^{+03}$	&4.95$\times$10$^{+02}$	 &7.04$\times$10$^{+03}$	 &4.16$\times$10$^{+03}$	 &2.96$\times$10$^{-09}$&SF\\
\\
$^{307}${121}& 11.64&	6.85$\times$10$^{-02}$&	6.74$\times$10$^{-03}$&5.94$\times$10$^{-02}$&	8.49$\times$10$^{-03}$&	 7.84$\times$10$^{-02}$&	 4.03$\times$10$^{-02}$&	 6.05$\times$10$^{+13}$&  $\alpha1$ \\
$^{303}${119}& 9.78	&2.01$\times$10$^{+03	}$&5.67$\times$10$^{+01	}$&1.37$\times$10$^{+03}$&	4.26$\times$10$^{+02}$&	 1.78$\times$10$^{+03}$&	 9.23$\times$10$^{+02}$&	 1.85$\times$10$^{+06}$&  $\alpha2$ \\
$^{299}${Ts}& 9.76	&5.23$\times$10$^{+02}$	&1.87$\times$10$^{+01}$	&3.49$\times$10$^{+02}$&	9.54$\times$10$^{+01}$&	 4.43$\times$10$^{+02}$&	 2.39$\times$10$^{+02}$&	 1.49$\times$10$^{+00}$&  SF      \\
$^{295}${Mc}& 9.78	&9.57$\times$10$^{+01}$	&4.48$\times$10$^{+00}$	&6.27$\times$10$^{+01}$&	1.48$\times$10$^{+01}$&	 7.85$\times$10$^{+01}$&	 4.40$\times$10$^{+01}$&	 2.61$\times$10$^{-05}$&  SF      \\
$^{291}${Nh}& 8.86	&1.64$\times$10$^{+04}$	&4.94$\times$10$^{+02}$	&9.63$\times$10$^{+03}$&	3.27$\times$10$^{+03}$&	 1.17$\times$10$^{+04}$&	 6.73$\times$10$^{+03}$&	 8.05$\times$10$^{-09}$&  SF    \\
$^{287}${Rg}& 8.85	&3.63$\times$10$^{+03}$	&1.44$\times$10$^{+02}$	&2.10$\times$10$^{+03}$&	6.30$\times$10$^{+02}$&	 2.55$\times$10$^{+03}$&	 1.51$\times$10$^{+03}$&	 3.60$\times$10$^{-11}$&  SF       \\
$^{283}${Mt}& 9.93	&3.85$\times$10$^{-01}$	&3.91$\times$10$^{-02}$	&2.40$\times$10$^{-01}$&	3.64$\times$10$^{-02}$&	 3.02$\times$10$^{-01}$&	 1.84$\times$10$^{-01}$&	 1.91$\times$10$^{-12}$& SF    \\
\\
$^{308}${121}& 12.00&2.11$\times$10$^{-02}$&	1.22$\times$10$^{-03}$	&2.14$\times$10$^{-02}$&	 1.02$\times$10$^{-03}$	&4.04$\times$10$^{-02}$	 &9.42$\times$10$^{-02}$	 &3.03$\times$10$^{+13}$& $\alpha1$\\
$^{304}${119}& 9.54	&2.52$\times$10$^{+04}$&	2.62$\times$10$^{+02}$	&2.69$\times$10$^{+04}$&	 2.61$\times$10$^{+03}$	&3.50$\times$10$^{+04}$	 &1.97$\times$10$^{+04}$	 &9.23$\times$10$^{+05}$&$\alpha2$\\
$^{300}${Ts}& 9.43	&1.21$\times$10$^{+04}$&	1.50$\times$10$^{+02}$	&1.26$\times$10$^{+04}$&	 1.12$\times$10$^{+03}$	&1.59$\times$10$^{+04}$	 &9.48$\times$10$^{+03}$	 &7.47$\times$10$^{-01}$&SF\\
$^{296}${Mc}& 9.49	&1.64$\times$10$^{+03}$&	2.79$\times$10$^{+01}$	&1.62$\times$10$^{+03}$&	 1.26$\times$10$^{+02}$	&2.10$\times$10$^{+03}$	 &1.49$\times$10$^{+03}$	 &1.31$\times$10$^{-05}$&SF\\
$^{292}${Nh}& 9.26	&1.81$\times$10$^{+03}$&	3.37$\times$10$^{+01}$	&1.76$\times$10$^{+03}$&	 1.33$\times$10$^{+02}$	&2.20$\times$10$^{+03}$	 &1.52$\times$10$^{+03}$	 &4.03$\times$10$^{-09}$&SF\\
$^{288}${Rg}& 8.48	&1.48$\times$10$^{+05}$&	2.03$\times$10$^{+03}$	&1.55$\times$10$^{+05}$&	 1.35$\times$10$^{+04}$	&1.61$\times$10$^{+05}$	 &7.06$\times$10$^{+04}$	 &1.80$\times$10$^{-11}$&SF\\
$^{284}${Mt}& 7.75	&1.40$\times$10$^{+07}$&	1.51$\times$10$^{+05}$	&1.64$\times$10$^{+07}$&	 1.63$\times$10$^{+06}$	&1.38$\times$10$^{+07}$	 &3.79$\times$10$^{+06}$	 &9.57$\times$10$^{-13}$&SF\\
\\
\hline
\hline
\end{tabular}}}
\end{table*}
\begin{table*}[htbp]
\caption{Comparison of Q$_{\alpha}$ values, $\alpha$-decay half-lives calculated by using modified Royer formula \cite{Akrawy2017} and possible mode of decay of Z = 121 $\alpha$-decay chains with available experimental data \cite{Oganessian2015}.}
\centering
\resizebox{0.6\textwidth}{!}{%
%\textbf{
{\begin{tabular}{l|l|l|l|l|l|l}
 \hline
 \hline
 \multicolumn{1}{l|}{Nuclei}&
 \multicolumn{2}{l|}{Q$_{\alpha}$(MeV)}&
 \multicolumn{2}{c|}{T$_{\alpha}${(1/2)(sec)}}&
 \multicolumn{2}{l}{Decay Mode}\\
 \cline{2-7}
 \multicolumn{1}{l|}{}&
 \multicolumn{1}{l|}{RMF}&
 \multicolumn{1}{l|}{Expt.}&
  \multicolumn{1}{l|}{RMF}&
       \multicolumn{1}{l|}{Expt.}&
        \multicolumn{1}{l|}{RMF}&
         \multicolumn{1}{l}{Expt.}\\
\hline
 $^{298}${121}&12.81&	&2.91$\times$10$^{-03}$&   &$\alpha1$ &\\
$^{294}${119}&11.29&	&9.74$\times$10$^{-01}$&   &$\alpha2$ &\\
$^{290}${Ts}&10.82&	    &3.04$\times$10$^{+00}$&   &$\alpha3$&\\
$^{286}${Mc}&11.87&     &4.93$\times$10$^{-03}$&   &$\alpha4$ &\\
$^{282}${Nh}&11.27&10.78$\pm$0.08&2.30$\times$10$^{-02}$&7.3$\times$10$^{-02}$&$\alpha5$ &$\alpha$ \\
$^{278}${Rg}&10.49&10.85$\pm$0.08&3.06$\times$10$^{-01}$&4.2$\times$10$^{-03}$&$\alpha6$&$\alpha$ \\
$^{274}${Mt}&9.85&10.20$\pm$1.10&2.68$\times$10$^{+00}$&4.4$\times$10$^{-01 }$&$\alpha7$/SF& $\alpha$\\
\hline
$^{299}${121}&12.42&  &8.14$\times$10$^{-04}$&  &$\alpha1$ & \\
$^{295}${119}&11.17&  &1.73$\times$10$^{-01}$&  &$\alpha2$ &\\
$^{291}${Ts}&10.72&	  &6.25$\times$10$^{-01}$&  &$\alpha3$&\\
$^{287}${Mc}&11.21&10.76$\pm$0.05&9.20$\times$10$^{-03}$&3.7$\times$10$^{-02}$&$\alpha4$ &$\alpha$  \\
$^{283}${Nh}&11.00&10.23$\pm$0.01&7.54$\times$10$^{-03}$&7.5$\times$10$^{-02}$&$\alpha5$ &$\alpha$   \\
$^{279}${Rg}&10.52&10.38$\pm$0.16&2.89$\times$10$^{-02}$&9.0$\times$10$^{-02}$&$\alpha6$&$\alpha$     \\
$^{275}${Mt}&10.90&10.48$\pm$0.01&8.67$\times$10$^{-04}$&2.0$\times$10$^{-02}$&$\alpha7$&$\alpha$\\
\hline
$^{300}${121}&12.76&   &3.42$\times$10$^{-03}$&    &$\alpha1$ &\\
$^{296}${119}&11.11&   &2.42$\times$10$^{+00}$&    &$\alpha2$ &\\
$^{292}${Ts}&10.93&	   &1.65$\times$10$^{+00}$&    &$\alpha3$&\\
$^{288}${Mc}&10.13&10.63$\pm$0.01&3.44$\times$10$^{+01}$&1.64$\times$10$^{-01}$&$\alpha4$&$\alpha$\\
$^{284}${Nh}&10.82&10.12$\pm$0.01&2.01$\times$10$^{-01}$&9.1$\times$10$^{-03}$&$\alpha5$&$\alpha$\\
$^{280}${Rg}&10.43&9.91$\pm$0.01 &4.10$\times$10$^{-01}$&4.6$\times$10$^{+00}$&SF&$\alpha$\\
$^{276}${Mt}&9.99&10.03$\pm$0.01 &1.18$\times$10$^{+00}$& 4.5$\times$10$^{-01}$&SF&$\alpha$  \\
\hline
$^{301}${121}& 12.33&	&1.25$\times$10$^{-03}$&   &$\alpha1$ &\\
$^{297}${119}& 10.86&	&1.02$\times$10$^{+00}$&   &$\alpha2$ &\\
$^{293}${Ts}& 10.91&11.32$\pm$0.05&1.87$\times$10$^{-01}$&2.2$\times$10$^{-02 }$  &$\alpha3$& $\alpha$\\
$^{289}${Mc}& 10.01&10.49$\pm$0.05&1.12$\times$10$^{+01}$&3.3$\times$10$^{-02}$  &$\alpha4$&$\alpha$ \\
$^{285}${Nh}& 10.30&10.01$\pm$0.04&1.19$\times$10$^{+00}$&4.2$\times$10$^{+00}$&$\alpha5$/SF&$\alpha$\\
$^{281}${Rg}& 10.37&9.41$\pm$0.05&6.83$\times$10$^{-02}$&1.7 $\times$10$^{+01}$  &SF      & SF \\
$^{277}${Mt}& 10.85&	&1.08$\times$10$^{-03}$&5.0$\times$10$^{-03}$   &SF      &SF\\
\hline
$^{302}${121}& 12.59&    &6.90$\times$10$^{-03}$&  &$\alpha1$ &\\
$^{298}${119}& 10.58&    &4.11$\times$10$^{+01}$&  &$\alpha2$ &\\
$^{294}${Ts}& 10.75&11.18$\pm$0.04&4.16$\times$10$^{+00}$& 5.1$\times$10$^{-02}$&$\alpha3$&$\alpha$\\
$^{290}${Mc}& 10.27&10.41$\pm$0.04&1.49$\times$10$^{+01}$&6.5$\times$10$^{-01}$ &$\alpha4$& $\alpha$ \\
$^{286}${Nh}& 9.33&9.79$\pm$0.05&1.04$\times$10$^{+03}$&9.5$\times$10$^{+00}$  &SF      &$\alpha$ \\
$^{282}${Rg}& 9.98&9.16$\pm$0.03&4.94$\times$10$^{+00}$&10.0$\times$10$^{+01}$  &SF      &$\alpha$ \\
$^{278}${Mt}& 9.93&9.58$\pm$0.03&1.56$\times$10$^{+00}$&4.5$\times$10$^{+00}$  &SF      &$\alpha$ \\
\hline
\hline
\end{tabular}}}
\end{table*}

\section{Conclusions}
We have employed relativistic mean-field (RMF) plus BCS approach and the Macroscopic-Microscopic approach with Nilson Strutinsky prescription
for an extensive and systematic study of even and odd isotopes of Z $=$ 121 (A $=$ 281 $-$ 380). We investigate ground state properties such as binding energy, separation energy, shell correction, deformation, shape, radii and charge density, and compared our results with the available results of FRDM which show good agreement with our calculations. A  complete trace of separation energy, shell correction, deformation and shape is presented which shows the strong evidences for magicity in N $=$ 228 and 164. We predict the central depletion in charge density for few isotopes of Z = 121 which is due to strong Coulomb repulsion due to a large number of protons. Depletion fraction is computed which is found to be stronger towards the neutron deficient side and decreases with increasing neutron number N which provide extra attractive force to balance the coloumb repulsion. \par We investigate the decay properties of a series of Z $=$ 121 isotopes (A $=$ 281 $-$ 380) and present our results on the modes of decay of $^{293-312}${121} and their decay products by comparing the alpha half-lives and the corresponding SF half-lives. We found that the nuclei with A $>$ 312 decay through spontaneous fission whereas those with A $<$ 312 decay through $\alpha$-decay. Nuclei with A = 293 - 302 are found with long $\alpha$-decay chains and hence the potential candidates for future experiments. This study may provide useful inputs for future experiments and search of new elements in superheavy region. However the scope to explore superheavy nuclei is boundless for theoretical as well as experimental research.

\section{Acknowledgement}
Authors would like to thank Prof. H. L. Yadav, BHU, India and Prof. L. S. Geng, Beihang University, China for their guidance and support. G.S. and M.A. acknowledge the support by SERB for YSS/2015/000952 and WOS-A schemes respectively.
%\clearpage
\newpage
\section{Appendix}%\vfill
To calculate $log_{10}T_{\alpha}$, we start with old, very famous and most frequently used expressions given by Viola and Seaborg (VSS) \cite{VSS1966} as
\begin{equation}
 log_{10}T_{\alpha}(sec) = (aZ + b){{Q_{\alpha}}^{-1/2}} + cZ + d + h_{log}
\end{equation}
where Z is the atomic number of parent nuclei and $h_{log}$ is the hindrance factor for nuclei (even and odd are denoted by 'e' and 'o', respectively) with the given values \cite{VSS1966}
\begin{table}[htbp]
\centering
%\resizebox{1.0
\begin{tabular}{|c|c|c|c|c|}
\hline
Nuclei (Z-N)&$e-e$&$o-e$&$e-o$&$o-o$\\
\hline
$h_{log}$&0.00&0.772&1.066&1.114\\
\hline
\end{tabular}
\end{table}
and the constants are a $=$ 1.66175, b $=$ -8.5166, c $=$ -0.20228, d $=$ -33.9069 taken from Ref. \cite{Sobiczewski1989}.\\
Another formula introduced by B. A. Brown \cite{Brown1992} is also used here which is given by:
 \begin{equation}
 log_{10}T_{\alpha}(sec) = 9.54\frac{(Z-2)^{0.6}}{\sqrt{Q_{\alpha}}} - 51.37
\end{equation}
where Z is the atomic number of parent nucleus.\\
In addition to this, we have also used the formula given by Sobiczewski in 2005 \cite{Sobiczewski2005}.
\begin{equation}
 log_{10}T_{\alpha}(sec) = aZ(Q_{\alpha}-\overline{E}_{i})^{-1/2} + bZ + c
\end{equation}
in which value of $\overline{E}_{i}$ is given by
\begin{table}[htbp]
%\centering
%\resizebox{1.0
\begin{tabular}{|c|c|c|c|c|}
\hline
Nuclei (Z-N)&$e-e$&$o-e$&$e-o$&$o-o$\\
\hline
&&&&\\[-1em]
$\overline{E}_{i}$&0.00&$\overline{E}_{p}$&$\overline{E}_{n}$&$\overline{E}_{p}$+$\overline{E}_{n}$\\
\hline
\end{tabular}
\end{table}

where $\overline{E}_{p}$ = 0.113 MeV and $\overline{E}_{n}$ = 0.171 MeV.\par
In 2009, universal decay law (UDL) is introduced by Qi and co-workers for $\alpha$ and cluster decay modes which is given as \cite{Qi2009}:
\begin{equation}
 log_{10}T_{\alpha}(sec) = aZ_{\alpha}Z_{d}\sqrt{\frac{A}{Q_{\alpha}}}+ b\sqrt{AZ_{\alpha}Z_{d}(A_{d}^{1/3} + A_{\alpha}^{1/3})} + c
\end{equation}
where A = $\frac{A_{d}A_{\alpha}}{A_{d}+A_{\alpha}}$ and the constants are a = 0.4314, b = -0.4087,and c = -25.7725. \par
From some of the latest formula, we have applied unified formula for ${\alpha}$-decay and cluster decay which is given as \cite{Ni2011}:
\begin{equation}
 log_{10}T_{\alpha}(sec) = a\sqrt{\mu}Z_{\alpha}Z_{d}Q_{\alpha}^{-1/2}+ b\sqrt{\mu}(Z_{\alpha}Z_{d})^{1/2}+c
\end{equation}
where a, b and c are constants a = 0.40, b = -1.31 and different values of c are
\begin{table}[htbp]
\centering
%\resizebox{1.0
\begin{tabular}{|c|c|c|c|c|}
\hline
Nuclei (Z-N)&$e-e$&$o-e$&$e-o$&$o-o$\\
\hline
c&-17.01&-16.40&-16.26&-15.85\\
\hline
\end{tabular}
\end{table}
and ${\mu}$ is the reduced mass.\\

Furthermore, we have also analyzed our result using most recent modified Royer formula given by Akrawy \textit{et al.} in 2017 \cite{Akrawy2017}.
\begin{equation}
 log_{10}T_{\alpha}(sec) = a + bA^{1/6}\sqrt{Z} + \frac{cZ}{\sqrt{Q_{\alpha}}}+ dI + eI^{2}
\end{equation}
where I = $\frac{N-Z}{A}$ and the constants a, b, c, d, and e are\\
\begin{table}[htbp]
%\caption{Equation 18}
\centering
\resizebox{0.45\textwidth}{!}{%
{\begin{tabular}{|c|c|c|c|c|c|}
\hline
 \multicolumn{1}{|c|}{Nuclei (Z-N)}&
 \multicolumn{1}{|c|}{a}&
 \multicolumn{1}{|c|}{b}&
 \multicolumn{1}{|c|}{c}&
 \multicolumn{1}{|c|}{d}&
 \multicolumn{1}{|c|}{e}\\
 \hline
 $e-e$&-27.837&-0.9420&1.5343&-5.7004&8.785\\
 $o-e$&-26.801&-1.1078&1.5585&14.8525&-30.523\\
 $e-o$&-28.225&-0.8629&1.5377&-21.145&53.890\\
 $o-o$&-23.635&-0.891&1.404&-12.4255&36.9005\\
\hline
\end{tabular}}}
\end{table}


\begin{thebibliography}{86}
\bibitem{Hofmann2000}S. Hofmann and G. Munzenberg, Rev. Mod. Phys. \textbf{72}, 733 (2000).
\bibitem{Hofmann2011}S. Hofmann, Radiochimica Acta International journal for chemical aspects of nuclear science and technology \textbf{99}, 405 (2011).
\bibitem{Morita2007}K. Morita, K. Morimoto, D. Kaji, T. Akiyama, S. ichi Goto, H. Haba, E. Ideguchi, K. Katori, H. Koura, H. Kikunaga, \textit{et al}., Journal of the Physical Society of Japan \textbf{76}, 045001 (2007).
\bibitem{Oganessian2010}Y.T. Oganessian, F.S. Abdullin, P.D. Bailey, D.E. Benker, M.E. Bennett, S.N. Dmitriev, J.G. Ezold, J.H. Hamilton, R.A. Henderson, M.G. Itkis \textit{et al}., Phys. Rev. Lett. \textbf{104}, 142502 (2010).
\bibitem{Oganessian2015}Yu. T. Oganessian and V.K. Utyonkov, Nucl. Phys. A \textbf{944}, 62 (2015).
\bibitem{Hamilton2013}J.H. Hamilton, S. Hofmann, and Y.T. Oganessian, Ann. Rev. Nucl. Part. Sci. \textbf{63}, 383 (2013).
\bibitem{Hofmann2016}S. Hofmann \textit{et al}., Eur. Phys. J. A \textbf{52}, 180 (2016).
\bibitem{Oganessian2009}Yu. T. Oganessian \textit{et al}., Phys. Rev. C \textbf{79}, 024603 (2009).
\bibitem{Buck1992}B. Buck, A.C. Merchant, and S.M. Perez, Phys. Rev. C \textbf{45}, 2247 (1992).
\bibitem{Poenaru1984}D.N. Poenaru, M. Ivascu, A. Sandulescu, and W. Greiner, Phys. Rev. C \textbf{32}, 572 (1985).
\bibitem{Basu2002}D.N. Basu, Phys. Lett. B \textbf{566}, 90 (2003).
\bibitem{Zhang2007}H.F. Zhang and G. Royer, Phys. Rev. C \textbf{76}, 047304 (2007).
\bibitem{Sharma2004}M.M. Sharma, A.R. Farhan, and G. Munzenberg, Phys. Rev. C \textbf{71}, 054310 (2005).
\bibitem{Pei2007}J.C. Pei, F.R. Xu, Z.J. Lin, and E.G. Zhao, Phys. Rev. C \textbf{76}, 044326 (2007).
\bibitem{Moller1994}P. Moller and J.R. Nix, Journal of Physics G: Nuclear and Particle Physics \textbf{20}, 1681 (1994).
\bibitem{Rutz1996}K. Rutz, M. Bender, T. Burvenich, T. Schilling, P.-G. Reinhard, J.A. Maruhn, and W. Greiner, Phys. Rev. C \textbf{56}, 238 (1997).
\bibitem{Cwiok1996}S. Cwiok, J. Dobaczewski, P.H. Heenen, P. Magierski,and W. Nazarewicz, Nucl. Phys. A \textbf{611}, 211 (1996).
\bibitem{Zhang2004}W. Zhang, J. Meng, S.Q. Zhang, L.-S. Geng, and H. Toki, Nucl. Phys. A \textbf{753}, 106 (2005).
\bibitem{Bao2015}X.J. Bao, S.Q. Guo, H.F. Zhang, Y.Z. Xing, J.M. Dong, and J.Q. Li, J. Phys. G \textbf{42}, 085101 (2015).
\bibitem{Wang2015}Y.Z. Wang, S.J. Wang, Z.Y. Hou, and J.Z. Gu, Phys. Rev. C \textbf{92}, 064301 (2015).
\bibitem{Niyti2015}Niyti, G. Sawhney, M.K. Sharma, and R.K. Gupta, Phys. Rev. C \textbf{91}, 054606 (2015).
\bibitem{Heenen2015}P.H. Heenen, J. Skalski, A. Staszczak, and D. Vretenar, Nucl. Phys. A \textbf{944}, 415 (2015).
\bibitem{Santhosh2016}K.P. Santhosh and C. Nithya, Phys. Rev. C \textbf{94}, 054621 (2016).
\bibitem{Budaca2016}A.I. Budaca, R. Budaca, and I. Silisteanu, Nucl. Phys. A \textbf{951}, 60 (2016).
\bibitem{Liu2017}J.-H. Liu, S.-Q. Guo, X.-J. Bao, and H.-F. Zhang, Chin. Phys. C \textbf{41}, 074106 (2017).
\bibitem{Zhang2017}Y.L. Zhang and Y.Z. Wang, Nucl. Phys. A \textbf{966}, 102 (2017).
\bibitem{Poenaru2017}D.N. Poenaru, H. Stcker, and R.A. Gherghescu Eur. Phys. J. A \textbf{54}, 14 (2018).
\bibitem{Manjunatha2016}H.C. Manjunatha, Int. J. Mod. Phys. E \textbf{25}, 1650100 (2016).
\bibitem{Mehta2015}M.S. Mehta, H. Kaur, B. Kumar, and S.K. Patra, Phys. Rev. C \textbf{92}, 054305 (2015).
\bibitem{Santhosh2016a}K.P. Santhosh and C. Nithya, Eur. Phys. J. A \textbf{52}, 371 (2016).
\bibitem{Santhosh2016xrd}K.P. Santhosh, B. Priyanka, and C. Nithya, Nucl. Phys. A \textbf{955}, 156 (2016).
\bibitem{Manjunatha2016bbm}H.C. Manjunatha, Int. J. Mod. Phys. E \textbf{25}, 1650074 (2016).
\bibitem{Manjunatha2016zia}H.C. Manjunatha, Nucl. Phys. A \textbf{945}, 42 (2016).
\bibitem{Rather2016}A.A. Rather, M. Ikram, A.A. Usmani, B. Kumar, and S.K. Patra, Eur. Phys. J. A \textbf{52}, 372 (2016).
\bibitem{Wu2003}W. Zhe-Ying, X. Fu-Rong, Z. En-Guang, and Z. Chun-Kai, Chinese Physics Letters \textbf{20}, 1702 (2003).
\bibitem{Adamian2009}G.G. Adamian, N.V. Antonenko, and V.V. Sargsyan, Phys. Rev. C \textbf{79}, 054608 (2009).
\bibitem{Biswal2014}S.K. Biswal, M. Bhuyan, S.K. Singh, and S.K. Patra, Int. J. Mod. Phys. E \textbf{23}, 1450017 (2014).
\bibitem{Berger2001}J.-F. Berger, L. Bitaud, J. Decharg, M. Girod, and K. Dietrich, Nuclear Physics A \textbf{685}, 1 (2001).
\bibitem{Decharge2003}J. Decharg, J.F. Berger, M. Girod, and K. Dietrich, Nucl. Phys. A \textbf{716}, 55 (2003).
\bibitem{Afanasjev2005}A.V. Afanasjev and S. Frauendorf, Phys. Rev. C \textbf{71}, 024308 (2005).
\bibitem{SinghSK2013}S.K. Singh, M. Ikram, and S.K. Patra, Int. J. Mod. Physics E \textbf{22}, 1350001 (2013).
\bibitem{Saxenaplb2018}G. Saxena, M. Kumawat, M. Kaushik, S.K. Jain, and M. Aggarwal, Communicated to Phys. Lett. B (2018).
\bibitem{Serot1984}B.D. Serot and J.D. Walecka, Adv. Nucl. Phys. \textbf{16}, 1 (1986).
\bibitem{Ring1996}P. Ring, Prog. Part. Nucl. Phys. \textbf{37}, 193 (1996).
\bibitem{Yadav2004}H.L. Yadav, M. Kaushik, and H. Toki, Int. J. Mod. Phys. E \textbf{13}, 647 (2004).
\bibitem{Saxena2017}G. Saxena, M. Kumawat, M. Kaushik, S.K. Jain, and M. Aggarwal, Phys. Lett. B \textbf{775}, 126 (2017).
\bibitem{Saxena2017hzo}G. Saxena, M. Kumawat, M. Kaushik, U.K. Singh, S.K. Jain, S.S. Singh, and M. Aggarwal, Int. J. Mod. Phys. E \textbf{26}, 1750072 (2017).
\bibitem{Aggarwal2010}M. Aggarwal, Phys. Lett. B \textbf{693}, 489 (2010).
\bibitem{Aggarwal2014}M. Aggarwal, Phys. Rev. C \textbf{89}, 024325 (2014).
\bibitem{Wang-Mass2017}M. Wang, G. Audi, F. Kondev, W. Huang, S. Naimi, and X. Xu, Chinese Physics C \textbf{41}, 030003 (2017).
\bibitem{Moller2012}P. Moller, A. Sierk, T. Ichikawa, and H. Sagawa, Atomic Data and Nuclear Data Tables \textbf{109-110}, 1 (2016).
\bibitem{Singh2013}D. Singh, G. Saxena, M. Kaushik, H.L. Yadav, and H. Toki, Int. J. Mod. Phys. E \textbf{21}, 9 (2012).
\bibitem{Geng2003}L.-S. Geng, H. Toki, S. Sugimoto, and J. Meng, Prog. Theor. Phys. \textbf{110}, 921 (2003).
\bibitem{Gambhir1989}Y.K. Gambhir, P. Ring, and A. Thimet, Annals Phys. \textbf{198}, 132 (1990).
\bibitem{Flocard1973}H. Flocard, P. Quentin, A.K. Kerman, and D. Vautherin, Nucl. Phys. A \textbf{203}, 433 (1973).
\bibitem{Geng2003wt}L.-S. Geng, H. Toki, A. Ozawa, and J. Meng, Nucl. Phys. A \textbf{730}, 80 (2004).
\bibitem{Dobaczewski1983}J. Dobaczewski, H. Flocard, and J. Treiner, Nucl. Phys. A \textbf{422}, 103 (1984).
\bibitem{Bertsch1991}G.F. Bertsch and H. Esbensen, Annals Phys. \textbf{209}, 327 (1991).
\bibitem{Dobaczewski1995}J. Dobaczewski, W. Nazarewicz, T.R. Werner, J.F. Berger, C.R. Chinn, and J. Decharge, Phys. Rev. C \textbf{53}, 2809 (1996).
\bibitem{Goriely2002}S. Goriely, M. Samyn, P.H. Heenen, J.M. Pearson, and F. Tondeur, Phys. Rev. C \textbf{66}, 024326 (2002).
\bibitem{Bender1999}M. Bender, K. Rutz, P.-G. Reinhard, and J.A. Maruhn, Eur. Phys. J. A \textbf{7}, 467 (2000).
\bibitem{Strutinsky1968}V.M. Strutinsky, Nucl. Phys. A \textbf{122}, 1 (1968).
\bibitem{BRACK1972}M. Brack, J. Damgaard, A.S. Jensen, H.C. Pauli, V.M. Strutinsky, and C.Y. Wongo, Rev. Mod. Phys. \textbf{44}, 320 (1972).
\bibitem{Nilsson1972}S.G. Nilsson and J. Damgaard, Physica Scripta \textbf{6}, 81 (1972).
\bibitem{Shanmugam1979}M.R.G. Shanmugam, P.R. Subramanian and V. Devanathan, The Present Status in Super-heavy Element Calculations, vol. \textbf{72} (1979).
\bibitem{EIS1976}J. Eisenberg and W. Greiner, Microscopic Theory of the Nucleus, Eisenberg: Nuclear theory (1972).
\bibitem{Seeger1975}P.A. Seeger and W.M. Howard, Nucl. Phys. A \textbf{238}, 491 (1975).
\bibitem{Lombard1976}R.J. Lombard, Phys. Lett. B \textbf{65}, 193 (1976).
\bibitem{Patra1999}S.K. Patra, C.-L. Wu, C.R. Praharaj, and R.K. Gupta, Nucl. Phys. A \textbf{651}, 117 (1999).
\bibitem{Wood1992}J.L. Wood, K. Heyde, W. Nazarewicz, M. Huyse, and P. van Duppen, Phys. Rept. \textbf{215}, 101 (1992).
\bibitem{Julin2001}R. Julin, K. Helariutta, and M. Muikku, Journal of Physics G: Nuclear and Particle Physics\textbf{ 27}, R109 (2001).
\bibitem{Li2016}J.J. Li, W.H. Long, J.L. Song, and Q. Zhao, Phys. Rev. C \textbf{93}, 054312 (2016).
\bibitem{Schuetrumpf2017}B. Schuetrumpf, W. Nazarewicz, and P.-G. Reinhard, Phys. Rev. C \textbf{96}, 024306 (2017).
\bibitem{Duguet2017} T. Duguet, V. Som, S. Lecluse, C. Barbieri, and P. Navrtil, Phys. Rev. C \textbf{95}, 034319 (2017).
\bibitem{Mutschler2016}A. Mutschler \textit{et al}., Nature Phys. \textbf{13}, 152 (2017).
\bibitem{Grasso2009}M. Grasso, L. Gaudefroy, E. Khan, T. Niksic, J. Piekarewicz, O. Sorlin, N.V. Giai, and D. Vretenar,  Phys. Rev. C \textbf{79}, 034318 (2009).
\bibitem{VSS1966}V. Viola and  G. Seaborg, Journal of Inorganic and Nuclear Chemistry \textbf{28}, 741 (1966).
\bibitem{Brown1992}B.A. Brown, Phys. Rev. C\textbf{ 46}, 811 (1992).
\bibitem{Sobiczewski2005}A. Parkhomenko and A. Sobiczewski, Acta Physica Polonica B \textbf{36}, 3115 (2005).
\bibitem{Qi2009}C. Qi, F.R. Xu, R.J. Liotta, and R. Wyss, Phys. Rev. Lett. \textbf{103}, 072501 (2009).
\bibitem{Ni2011}D. Ni and Z. Ren, Rom. Journ. Phys. \textbf{57}, 407 (2012).
\bibitem{Akrawy2017}D.T. Akrawy and D.N. Poenaru, J. Phys. G \textbf{44}, 105105 (2017).
\bibitem{Xu2008}C. Xu, Z. Ren, and Y. Guo, Phys. Rev. C \textbf{78}, 044329 (2008).
\bibitem{Sobiczewski1989}A. Sobiczewski, Z. Patyk, and S. wiok, Phys. Lett. B \textbf{224}, 1 (1989).
\end{thebibliography}
\end{document}